\documentclass[10pt,conference,compsoc]{IEEEtran}

\newif\ifFinal
\Finaltrue  

\newif\ifExtended
\Extendedtrue

\newenvironment{DIFnomarkup}{}{}

\usepackage{pifont}
\usepackage{cite}
\usepackage{color}
\usepackage{array}
\usepackage{mathpartir}
\usepackage{xifthen}

\definecolor{lcol}{rgb}{0,0,0.5}

\usepackage[pdftex,hyperindex=false,bookmarks=false,
	        colorlinks=true,linkcolor=lcol,citecolor=lcol,
            filecolor=lcol,pagecolor=lcol,urlcolor=lcol]{hyperref}

\usepackage{url}
\usepackage{xspace}
\usepackage{listings}
\usepackage{amssymb}
\usepackage{amsmath}
\usepackage{amsthm}

\usepackage[pdftex]{graphicx}

\usepackage{tikz}
\usepackage{subcaption}

\usepackage{color}

\usepackage{enumitem}
\setitemize{noitemsep,nosep,leftmargin=3mm}

\newcommand{\subparagraph}{}
\usepackage[compact]{titlesec}  

\newcommand{\nbc}[3]{
 {\colorbox{#3}{\bfseries\sffamily\scriptsize\textcolor{white}{#1}}}
 {\textcolor{#3}{\sf\small$\blacktriangleright$\textit{#2}$\blacktriangleleft$}}}

\definecolor{dscolor}{rgb}{0.1,0.1,0.9}
\newcommand\ds[1]{\nbc{DS}{#1}{dscolor}}

\newtheorem{definition}{Definition}[subsection]

\newtheorem{theorem}{Theorem}[subsection]

\newcounter{linenumbercounter}
\definecolor{anncol}{rgb}{0,0.5,0}
\newlength{\codeindentlength}
\newcommand{\increaseindent}{\addtolength{\codeindentlength}{1em}}
\newcommand{\decreaseindent}{\addtolength{\codeindentlength}{-1em}}
\newcommand{\indentcode}{\hspace*{\codeindentlength}}

\newcommand{\printcodelinenumber}{{\scriptsize\arabic{linenumbercounter}}\stepcounter{linenumbercounter}}
\newcommand{\codestart}{\setcounter{linenumbercounter}{1}\printcodelinenumber\,}

\newcommand{\codenewline}{\\\printcodelinenumber\indentcode}

\newenvironment{examplecode}{\codestart}{}

\newcommand{\ann}[1]{\color{anncol}\{#1\}\normalcolor}
\newcommand{\kw}[1]{\textbf{#1}}
\newcommand{\Skip}{\kw{stop}}
\newcommand{\Seq}[2]{#1;\codenewline#2}
\newcommand{\Nop}[1]{\ann{#1} \kw{nop}}
\newcommand{\Acquire}[2]{\ann{#1} \kw{acquire}(#2)}
\newcommand{\Release}[2]{\ann{#1} \kw{release}(#2)}
\newcommand{\Assign}[3]{\ann{#1} #2\ensuremath{{}\mathbin{:=}{}}#3}
\newcommand{\dassign}{\widehat{:=}}
\newcommand{\DAssign}[3]{\ann{#1} #2\ensuremath{{}\mathbin{\dassign}{}}#3}
\newcommand{\Input}[3]{\ann{#1} #2\ensuremath{{}\mathbin{\leftarrow}{}}#3}
\newcommand{\Output}[3]{\ann{#1} #2\ensuremath{{}\mathbin{!}{}}#3}
\newcommand{\doutput}{\widehat{!}}
\newcommand{\DOutput}[3]{\ann{#1} #2\ensuremath{{}\mathbin{\doutput}{}}#3}
\newcommand{\ITE}[4]{\ann{#1} \kw{if}\ #2\increaseindent\codenewline#3\decreaseindent\codenewline\kw{else}\increaseindent\codenewline#4\decreaseindent\codenewline\kw{endif}}
\newcommand{\ITENoEndIf}[4]{\ann{#1} \kw{if}\ #2\increaseindent\codenewline#3\decreaseindent\codenewline\kw{else}\increaseindent\codenewline#4\decreaseindent}
\newcommand{\ITNoElse}[3]{\ann{#1} \kw{if}\ #2\increaseindent\codenewline#3\decreaseindent\codenewline\kw{endif}}

\newcommand{\var}[1]{\ensuremath{\mathit{#1}}\xspace}
\newcommand{\const}[1]{\textsf{#1}\xspace}

\newcommand{\true}{\const{true}}
\newcommand{\false}{\const{false}}

\newcommand{\ITEg}[4]{\ann{#1} \kw{if}\ #2 #3 \kw{else} #4 \kw{endif}}
\newcommand{\Whileg}[4]{\ann{#1} \kw{while}\ #2\ \kw{inv} \ann{#3} \kw{do} #4}
\newcommand{\Seqg}[2]{#1;\ #2}
\newcommand{\ITNoElseg}[3]{\ann{#1} \kw{if}\ #2 #3 \kw{endif}}

\newcounter{anncounter}
\newcommand{\nextann}{$A_{\arabic{anncounter}}$\stepcounter{anncounter}}

\newcommand{\varl}{\ell}
\newcommand{\buf}{\var{buf}}
\newcommand{\valid}{\var{valid}}
\newcommand{\GHOST}{\var{GHOST}}
\newcommand{\lbuf}{\bot\var{buf}}
\newcommand{\hbuf}{\top\var{buf}}
\newcommand{\ldecl}{\var{d}}
\newcommand{\CHECKSIG}{\var{CK}}
\newcommand{\inmode}{\var{inmode}}
\newcommand{\outmode}{\var{outmode}}
\newcommand{\useranswer}{\var{answer}}

\newcommand{\HasType}[2]{#1 \vdash #2}
\renewcommand{\L}{\mathcal{L}}
\newcommand{\D}{\mathcal{D}}
\newcommand{\E}{\mathcal{E}}

\newcommand{\R}{\mathcal{R}}
\newcommand{\explevel}[3]{\mathit{sensitivity}(#2,#3,#1)}  
\newcommand{\bisimilar}[3]{#2 \mathrel{\stackrel{#1}{\sim}} #3}

\newcommand{\lsrc}{\mathit{lvl}_{\mathit{src}}}
\newcommand{\ldst}{\mathit{lvl}_{\mathit{dst}}}
\newcommand{\lvl}{\mathit{lvl}}
\newcommand{\Alvl}{\mathit{lvl}_\mathsf{A}} 

\newcommand{\env}[1][]{\mathit{env}\ifthenelse{\equal{#1}{}}{}{_{#1}}}
\newcommand{\trace}[1][]{\mathit{tr}\ifthenelse{\equal{#1}{}}{}{_{#1}}}
\newcommand{\locks}[1][]{\mathit{locks}\ifthenelse{\equal{#1}{}}{}{_{#1}}}
\newcommand{\mem}[1][]{\mathit{mem}\ifthenelse{\equal{#1}{}}{}{_{#1}}}

\newcommand{\ls}[1][]{\mathit{ls}\ifthenelse{\equal{#1}{}}{}{_{#1}}}

\newcommand{\sched}{\mathit{sched}}

\newcommand{\vs}{\mathit{vs}} 
\newcommand{\xs}{\mathit{xs}} 

\newcommand{\IfThenElse}[3]{\mathrm{\bf if}\ #1\ \mathrm{\bf then}\ #2\ \mathrm{\bf else}\ #3}
\newcommand{\take}{\mathit{take}}
\newcommand{\avg}{\mathit{avg}}
\newcommand{\rev}{\mathit{rev}}
\newcommand{\length}{\mathit{len}}
\newcommand{\last}{\mathit{last}}
\newcommand{\union}{\mathrel{\cup}}
\newcommand{\sigmainit}{\sigma_{\mathsf{init}}}
\newcommand{\lastinput}[2]{\mathit{lastinput}(#1,#2)}
\newcommand{\lastoutput}[2]{\mathit{lastoutput}(#1,#2)}
\newcommand{\levelinputs}[2]{\mathit{inputs}(#1,#2)}

\newcommand{\ievent}[2]{{\bf in}\langle#1,#2\rangle}
\newcommand{\oevent}[3]{{\bf out}\langle#1,#2,#3\rangle}
\newcommand{\devent}[3]{{\bf d}\langle#1,#2,#3\rangle}
\newcommand{\visible}[1]{\mathit{visible}_{#1}}
\newcommand{\filter}[2]{#2 \mathrel{\upharpoonright} {#1}}
\newcommand{\trequiv}[3]{#2 \stackrel{#1}{\approx} #3}
\newcommand{\envequiv}[3]{#2 \stackrel{#1}{\approx} #3}
\newcommand{\memequiv}[3]{#2 \stackrel{#1}{\approx} #3}
\newcommand{\gequiv}[3]{#2 \stackrel{#1}{\approx} #3}

\newcommand{\stepsecure}[5]{\mathit{stepsec}_{#1}(#2, #3, #4, #5)}

\newcommand{\uncertainty}[5]{\mathit{uncertainty}_{#1}(#2,#3,#4,#5)}
\newcommand{\eventsecure}[8]{\mathit{esec}_{#1}((#2,#3,#4),(#5,#6,#7),#8)}
\newcommand{\syssecure}[2]{\mathit{syssec}_{#1}(#2)}

\newcommand{\drsecure}[2]{\mathit{drsec}_{#1}(#2)}
\newcommand{\envdrequiv}[4]{#3 \stackrel{#1,#2}{\approx} #4}
\newcommand{\gdrequiv}[4]{#3 \stackrel{#1,#2}{\approx} #4}

\newcommand{\inputcount}{\mathit{cnt}}
\newcommand{\inputsum}{\mathit{sum}}
\newcommand{\readbuf}{\mathit{buf}}
\newcommand{\mininputs}{\mathit{min}}

\newcommand{\lstep}[2]{#1 \leadsto #2}

\newcommand{\eval}[2]{\lfloor#2\rfloor_{#1}}

\newcommand{\gstate}[1]{(\env[#1],\mem[#1],\locks[#1],\trace[#1])}
\newcommand{\gstep}[2]{#1 \rightarrow #2} 
\newcommand{\gstepstar}[2]{#1 \rightarrow^* #2} 

\newcommand{\annsat}[1]{\vDash #1}
\newcommand{\annsatLoc}[2]{#1 \vDash #2} 

\newcommand{\circled}[1]{\ding{\numexpr#1+191\relax}}

\newsavebox{\mybox}
\newenvironment{adversaryexpectation}{%
  \begin{lrbox}{\mybox}\begin{minipage}{0.94\columnwidth}{\bf Adversary Expectation:}}{%
  \end{minipage}\end{lrbox}\ \\\ \\\noindent\fbox{\usebox{\mybox}}\medskip \\}

\newenvironment{domainhypothesis}{%
  \begin{lrbox}{\mybox}\begin{minipage}{0.94\columnwidth}{\bf Domain Hypothesis}}{%
  \end{minipage}\end{lrbox}\ \linebreak\ \linebreak\noindent\fbox{\usebox{\mybox}}\medskip \\}

\newenvironment{domainhypothesisnobreak}{%
  \begin{lrbox}{\mybox}\begin{minipage}{0.94\columnwidth}{\bf Domain Hypothesis}}{%
  \end{minipage}\end{lrbox}\noindent\fbox{\usebox{\mybox}}\medskip \\}

\newenvironment{securityguarantee}{%
  \begin{lrbox}{\mybox}\begin{minipage}{0.94\columnwidth}{\bf Security Guarantee:}}{%
  \end{minipage}\end{lrbox}\ \\\ \\\noindent\fbox{\usebox{\mybox}}\medskip \\}

\newcommand{\OurSystem}{\textsc{Veronica}\xspace}
\newcommand{\Covern}{\textsc{Covern}\xspace}
\newcommand{\SecCSL}{\textsc{SecCSL}\xspace}

\newcommand{\pkgurl}{\cite{theories}}

\newcommand{\Title}{\OurSystem: Expressive and Precise Concurrent Information Flow Security\ifExtended\\(Extended Version with Technical Appendices)\fi}

\title{\Title}

\ifFinal 
\author{\IEEEauthorblockN{%
Daniel Schoepe\IEEEauthorrefmark{1},
Toby Murray\IEEEauthorrefmark{2} and
Andrei Sabelfeld\IEEEauthorrefmark{1}}
\IEEEauthorblockA{\IEEEauthorrefmark{1}Chalmers University of Technology \qquad
                  \IEEEauthorrefmark{2}University of Melbourne and Data61}
}
\fi

\begin{document}

\maketitle

\begin{abstract}
Methods for proving that concurrent software
does not leak its secrets has remained an active topic of research for at
least the past four decades.
Despite an impressive array of work, the present situation
remains highly unsatisfactory. With contemporary compositional
proof methods one is forced
to choose between \emph{expressiveness} (the ability to reason about a wide
variety of security policies), on the one hand, and
\emph{precision} (the ability to reason about complex thread interactions
and program behaviours), on the other. Achieving both is essential and,
we argue, requires a new style of compositional reasoning.

We present \OurSystem, 
the first program logic
for proving concurrent programs information flow secure that supports
compositional, high-precision reasoning about 
a wide range of security policies and program behaviours (e.g.\ expressive declassification, value-dependent classification, secret-dependent branching).
Just as importantly, \OurSystem  embodies
a new approach for engineering such logics
that can be re-used elsewhere, called
\emph{decoupled functional correctness} (DFC). DFC
leads to a 
simple and clean logic, even while achieving this unprecedented combination of
features. 
 We demonstrate the virtues and versatility of
 \OurSystem by verifying a range of example programs, beyond
 the reach of prior methods.

\end{abstract}

\section{Introduction}

%

Software guards our most precious secrets. More often than not, software
systems are built as a collection of concurrently executing threads of execution that
cooperate to process data. In doing so, these threads
collectively implement security policies
in which the sensitivity of the data being
processed is often data-dependent~\cite{Yang_HASFC_16,Zheng_Myers_07,Swamy_CC_10,Swamy_CCFSBY_11,Lourenco_Caires_15,Zhang_WSM_15,Amtoft_BB_06,Murray_MBGK_12,Nanevski_BG_11,Murray_SE_18},
and the rules about to whom it can
be disclosed and under what conditions can be non-trivial~\cite{Broberg_Sands_10,vanDelft_HS_15,Broberg_vDS_15,Askarov_Chong_12,Zhang_11,Eggert_vanderMeyden_17}.
The presence of concurrency greatly complicates reasoning, since a thread
that behaves securely when run in isolation can be woefully insecure in the
presence of interference from others~\cite{McCullough_87,Smith_Volpano_98,Murray_SPR_16,Murray_SE_18}
or due to scheduling~\cite{Volpano_Smith_99,Sabelfeld_Sands_00}.

For these reasons, being able to formally prove that concurrent software
does not leak its secrets (to the wrong places at the wrong times) has
been an active and open topic of research for at least the past four
decades~\cite{Andrews_Reitman_78,Andrews_Reitman_80}.
Despite an impressive array of work over that time, the present situation
remains highly unsatisfactory. With contemporary proof methods one is forced
to choose between \emph{expressiveness} (e.g.~\cite{Mantel_Reinhard_07,Bossi_PR_07,Lux_MP_12,Bauereiss_GPR_17}), on the one hand, and
\emph{precision} (e.g.~\cite{Ernst_Murray_19,Murray_SE_18,Karbyshev+:POST18,Murray_SPR_16,Mantel_SS_11,Neilson_NL_15,Nielson_Nielson_17,Li_MT_17}), on the other.

By expressiveness, we mean the ability to reason about the enforcement
of a wide variety of security policies and classes thereof,
such as state-dependent secure
declassification and data-dependent sensitivity.
It is well established that, beyond simple noninterference~\cite{Goguen_Meseguer_82}
(``secret data should never be revealed in public outputs''), there is no
universal solution to specifying information flow policies~\cite{Broberg_vDS_15}, and
that different applications
might have different
interpretations on what adherence to a particular policy means.


By precision, we mean the ability to reason about complex thread interactions
and program behaviours. This includes not just
program behaviours like secret-dependent branching that are beyond the scope
of many existing proof methods
(e.g.~\cite{Murray_SPR_16,Murray_SE_18,Ernst_Murray_19}).
Moreover, precision is aided by reasoning about each thread
under local assumptions that it makes about the behaviour of the others~\cite{Jones:phd,Mantel_SS_11}.
For instance~\cite{Murray_SE_18},
suppose
thread~$B$ receives data from thread~$A$, by acquiring a lock on a shared
buffer and then checking the buffer contents. Thread~$B$ relies on thread~$A$
having appropriately labelled the buffer to indicate the (data-dependent)
sensitivity of the
data it contains and, while thread~$B$ holds the lock, it relies on all other threads
to avoid modifying the buffer (to preserve the correctness of the
sensitivity label). Precise reasoning here should take account of
these kinds of assumptions when reasoning about thread~$B$ and,
correspondingly, should prove that they are adhered to 
when reasoning about thread~$A$.

Besides expressiveness and precision, another useful property
for a proof method to have is \emph{compositionality}.
We say that a proof method
is \emph{compositional}~\cite{McLean_94,Zakinthinos_Lee_97,Mantel_02,Rafnsson_Sabelfeld_14} when it can be used to establish the security of the entire
concurrent program by using it to prove each thread secure separately.

So far it has remained an open problem of how to design a proof method
(e.g.\ a security type system~\cite{Volpano_IS_96} or program logic~\cite{Amtoft_Banerjee_04})
that is (a)~compositional, (b)~supports proving a general enough definition
of security to encode a variety
of security policies, and (c)~supports precise reasoning.
%
%
We argue that achieving all three together
requires a
new style of program logic for information flow security.

In this paper, we present \OurSystem. \OurSystem is, to our knowledge,
the first compositional program logic
for proving concurrent programs information flow secure that supports
high-precision reasoning about
a wide range of security policies and program behaviours (e.g.\ expressive declassification, value-dependent classification, secret-dependent branching).
Just as importantly, \OurSystem  embodies
a new approach for engineering such logics
that can be re-used elsewhere. This approach we call
\emph{decoupled functional correctness} (DFC), which we have found
leads to a simple and
clean logic, even while achieving this unprecedented combination of
features. Precision is supported by reasoning about a program's
functional properties. However, the key insight of DFC is that this
reasoning can and should be separated from reasoning about its
information flow security. 
As we explain, DFC exploits
compositional functional correctness as a common means to unify together
reasoning about
various security concerns.

We provide an overview of \OurSystem in \autoref{sec:overview}.
\autoref{sec:security-property} then describes the general security property
that it enforces, and so formally defines the threat model. \autoref{sec:lang} describes the programming
language over which \OurSystem has been developed. \autoref{sec:logic} then
describes the \OurSystem logic, whose virtues are further demonstrated
in \autoref{sec:examples}. \autoref{sec:related-work} considers
related work before \autoref{sec:concl} concludes.

All results in this paper have been mechanised in
the interactive theorem prover Isabelle/HOL~\cite{Nipkow_PW:Isabelle}.
Our Isabelle formalisation is available online~\pkgurl.

\section{An Overview of \OurSystem}\label{sec:overview}


\subsection{Decoupling Functional Correctness}\label{sec:overview-dfc}

\newlength{\subfigslength}
\setlength{\subfigslength}{5cm}

\newlength{\figcellwidth}
\setlength{\figcellwidth}{\subfigslength}
\addtolength{\figcellwidth}{1cm}

\begin{figure*}
  \begin{center}
    \begin{tabular}{m{\figcellwidth}m{\figcellwidth}m{\figcellwidth}}
      \begin{tabular}{c}
      \begin{subfigure}[b]{\subfigslength}
\begin{tikzpicture}[scale=0.7]

\node at (-1,-0.5) {$\bot$};
\node at (-1,2.5) {$\top$};

\draw [->] (-0.75,-0.25) -- (0,0.75);
\draw [->] (-0.75,2.25) -- (0,1.25);

\node at (0.5,1) {$\buf$};

\node at (2.5,-0.5) {$\lbuf$};
\node at (2.5,2.5) {$\hbuf$};

\draw [<-] (1.75,-0.25) -- (1,0.75);
\draw [<-] (1.75,2.25) -- (1,1.25);

\draw [->] (3.25,2.5) -- (4.25,2.5);
\draw [->] (3.25,-0.5) -- (4.25,-0.5);

\node at (4.5,-0.5) {$\bot$};
\node at (4.5,2.5) {$\top$};

\draw [densely dotted,->] (2.5, 2.25) -- (2.5,-0.25);
\end{tikzpicture}    
    \caption{Data flows. Dotted lines denote declassification.\label{fig:example-arch}}
      \end{subfigure}
      \bigskip \\
  \begin{subfigure}[b]{\subfigslength}
    \begin{examplecode}
      \Output{\nextann}{$\top$}{$\hbuf$}
    \end{examplecode}
    \caption{Outputting $\top$ data.\label{fig:example-top}}
  \end{subfigure}
  \end{tabular}
      &
      \begin{tabular}{l}
      \begin{subfigure}[b]{\subfigslength}
    \begin{examplecode}
      \Seq{\Acquire{\nextann}{$\varl$}}{\Seq{\ITE{\nextann}{$\inmode$ = 0}{\Input{\nextann}{$\buf$}{$\bot$}}{\Input{\nextann}{$\buf$}{$\top$}}}}{\Seq{\Assign{\nextann}{$\valid$}{1}}{\Release{\nextann}{$\varl$}}}
    \end{examplecode}
    \caption{Reading data into a shared buffer.\label{fig:example-mux}}
      \end{subfigure}
      \bigskip \\
  \begin{subfigure}[b]{\subfigslength}
    \begin{examplecode}
      \Output{\nextann}{$\bot$}{$\lbuf$}
    \end{examplecode}
    \caption{Outputting $\bot$ data.\label{fig:example-bot}}
  \end{subfigure}
  \end{tabular}      
      &
  \begin{subfigure}[b]{\subfigslength}
    \begin{examplecode}
      \Seq{\Acquire{\nextann}{$\varl$}}
          {\Seq{\ITNoElse{\nextann}{$\valid = 1$}{\ITE{\nextann}{$\outmode$ = 0}
                    {\Assign{\nextann}{$\lbuf$}{$\buf$}}
                    {\Seq{\Assign{\nextann}{$\hbuf$}{$\buf$}}
                         {\Seq{\DAssign{\nextann}{$\ldecl$}{\CHECKSIG($\hbuf$)}}
                              {\ITNoElse{\nextann}{$\ldecl$ = 0}
                                        {\DAssign{\nextann}{$\lbuf$}{$\hbuf$}}
                              }
                         }
                    }}
               }{\Release{\nextann}{$\varl$}}}
    \end{examplecode}
    \caption{Copying and declassifying data.\label{fig:example-demux}}
  \end{subfigure}
    \end{tabular}
  \caption{Co-operative Use of a Shared Buffer\label{fig:example}. Green
    $\ann{A_i}$ are functional correctness annotations (whose contents we omit).}
  \end{center}
\end{figure*}

\autoref{fig:example-arch} depicts the data-flow architecture for a very simple,
yet illustrative, example system. This example is inspired by a real world
security-critical shared-memory concurrent program~\cite{Murray_SE_18}.
This example purposefully avoids
some of \OurSystem's features (e.g.\ secret-dependent
branching and runtime state-dependent declassification policies), which
we will meet later in \autoref{sec:examples}. Verifying it requires highly
precise reasoning, and the security policy it enforces involves both
data-dependent sensitivity and delimited release style
declassification~\cite{Sabelfeld_Myers_03}, features that until now have
never been reconciled before.

The system comprises four threads, whose code appears
in \autoref{fig:example} (simplified a little for presentation).
The four threads make use of a shared buffer~\buf protected by a
lock~$\varl$, which also protects the shared flag variable~$\valid$.
The top-middle thread (\autoref{fig:example-mux}) copies data into
the shared buffer, from one of two input/output (IO) \emph{channels}:~$\bot$
(a public channel whose contents is visible
to the attacker) and~$\top$ (a private channel,
not visible to the attacker). The right-top thread
(\autoref{fig:example-demux})
reads data from the shared
buffer~\buf and copies it to an appropriate output buffer (either $\lbuf$ for $\bot$ data or
$\hbuf$ for $\top$ data) for further processing by the remaining two output threads. 

Each of the bottom threads 
outputs from its respective output buffer to its
respective channel; one for~$\top$ data (\autoref{fig:example-top}) and the
other for~$\bot$ data (\autoref{fig:example-bot}).

The decision of the top-middle thread (\autoref{fig:example-mux}, line~2), whether to input from the $\bot$ channel or
the $\top$ one, is dictated by the shared variable~$\inmode$.
The $\valid$ variable (initially zero) is set to 1 by the top-middle thread once it has
filled the~$\buf$ variable, and is then tested by the top-right thread
(\autoref{fig:example-demux}, line~2) to ensure it doesn't consume data
from~$\buf$ before the top-middle thread has written to~$\buf$.

The
top-right thread's decision (\autoref{fig:example-demux}, line~3) about which output buffer it should copy the data in~$\buf$ to is dictated
by the~$\outmode$ variable. When~$\outmode$ indicates that the $\top$ buffer~$\hbuf$
should be used, the top-right thread additionally performs a
signature check (via the~\CHECKSIG function, lines 7--8) on the data to decide if it is safe
to declassify and copy additionally to the $\lbuf$ output buffer. This concurrent
program implements a \emph{delimited release}~\cite{Sabelfeld_Myers_03} style declassification
policy, which states that $\top$ data that passes the signature check,
plus the results of the signature check itself for all $\top$ data,
are safe to
declassify. The language of \OurSystem includes the \emph{declassifying
  assignment} command~$\dassign$ and the \emph{declassifying output}
command~$\doutput$.  Besides delimited release style declassification policies,
we we will see later in the examples of \autoref{sec:examples} that
our security condition also supports stateful declassification policies.

Clearly, if $\inmode$ and $\outmode$ disagree, the concurrent
execution of the threads might behave insecurely (e.g.\ the top-middle thread
might place private $\top$ data into $\buf$, which the top-right thread then copies
to~$\lbuf$ and is subsequently output on the
public channel~$\bot$).
Therefore, the security of this concurrent program rests on the shared data
\emph{invariant}
that~$\inmode$ and~$\outmode$ agree (whenever lock~$\varl$ is acquired and released).
This is a functional correctness property.
There are a number of other such functional
properties, somewhat more implicit, on which
the system's security relies, e.g.\ that neither thread will modify~$\buf$
unless they hold the lock~$\varl$, and likewise for~$\inmode$ and~$\outmode$, plus that
only one thread can hold the lock~$\varl$ at a time.

Similarly, the security of the declassification actions performed by the top-right
thread rests on the fact that it only declassifies after successfully performing
the signature check, in accordance with the
delimited release policy.

Thus one cannot reason about the security of this concurrent program in the absence
of functional correctness. However, one of the fundamental insights of
\OurSystem is that
functional correctness reasoning should be \emph{decoupled} from security reasoning.
This is in contrast
to many recent logics for concurrent information flow security, notably~\cite{Ernst_Murray_19},
the \Covern logic of~\cite{Murray_SE_18} and its antecedents~\cite{Murray_SPR_16,Mantel_SS_11} as well
as~\cite{Neilson_NL_15,Nielson_Nielson_17} plus many prior logics for
sequential programs~\cite{Zheng_Myers_07,Nanevski_BG_11,Banerjee_NR_08,Murray_MBGK_12,Khakpour_SD_13,Lourenco_Caires_15} and
hardware designs~\cite{Ferraiuolo_HMS_17}.


%
%

\OurSystem decouples functional correctness reasoning from security reasoning
by performing the latter over programs that carry
\emph{functional correctness annotations}~$\ann{A_i}$ on each program statement~$s_i$.
Thus program statements are of the form~$\ann{A_i}~s_i$. Here, $\ann{A_i}$ should be
thought of as akin to a Hoare logic precondition~\cite{Hoare_69}. It states conditions that are
known to be true whenever statement~$s_i$ is executed in the \emph{concurrent}
program.
We call this resulting approach \emph{decoupled functional correctness} (DFC).

The contents of each of the annotations in \autoref{fig:example-mux} and
\autoref{fig:example-demux} have been omitted in the interests of brevity (they can
be found in our Isabelle formalisation),
and simply replaced by identifiers~$\ann{A_i}$.

For verifying the security of the top-right
thread~(\autoref{fig:example-demux}), annotations $\ann{A_{11}}$ through $\ann{A_{15}}$ are most important: $\ann{A_{11}}$ would imply that \buf holds an input read from
channel~$\bot$ (justifying why copying its contents to the $\bot$ variable~$\lbuf$ is secure),
and~$\ann{A_{12}}$ would imply likewise for channel~$\top$.
$\ann{A_{13}}$ would imply that $\hbuf$ holds $\top$ data and
$\ann{A_{14}}$ that $d$ holds the result of the signature check. Finally,
$\ann{A_{15}}$ implies that the signature check passed, justifying why
the declassifying assignment to $\lbuf$ is secure.

The other annotations encode functional correctness information
needed to justify the validity of the aforementioned annotations.
For instance,
annotation~$\ann{A_2}$ in \autoref{fig:example-mux} implies
that the thread holds the lock~$\varl$; $\ann{A_3}$ that
$\inmode$ is zero, while $\ann{A_4}$ the opposite. Annotation~$\ann{A_5}$
on the
other hand tracks information about the contents of~$\buf$, namely if $\inmode$ is zero
then \buf holds the last input read from channel~$\bot$, and it holds the last
input read from channel~$\top$ otherwise\footnote{$\ann{A_5}$ effectively encodes
$\buf$'s (state-dependent) sensitivity, and takes the place of dependent
  security types and labels from prior systems.}.

Thus the annotations~$\ann{A_i}$ afford highly precise reasoning about
the security of each thread, while decoupling the
functional correctness reasoning.

The idea of using annotations~$\ann{A_i}$ we repurpose from the
Owicki-Gries proof technique~\cite{Owicki_Gries_76} for concurrent programs.
Indeed, there exist 
a range of standard techniques for inferring and proving the soundness of such
annotations (i.e.\ for carrying out the functional correctness reasoning),
from the past 40 years of research on concurrent program verification. \OurSystem integrates multiple such
techniques in the Isabelle/HOL theorem prover, each of which has been proved sound
from first principles, thereby ensuring the correctness of its foundations.
As we explain later,  external program verifiers may also
be used to verify functional correctness, giving up \OurSystem's foundational
guarantees in exchange for ease of verification.

Given a correctly annotated program, \OurSystem then uses the
information encoded in the annotations to prove expressive security
policies, as we outline in the next section.

\subsection{Compositional Enforcement}\label{sec:overview-comp}

How can we prove that the concurrent program of \autoref{fig:example} doesn't
violate information flow security, i.e.\ that no $\top$ data is
leaked, unless it has been declassified in accordance with the
delimited release policy?

Doing so in general benefits from having
a compositional reasoning method, namely one that
reasons over each of the program's threads separately to
deduce that the concurrent execution of those threads is
secure.

Compositional methods for proving information flow properties of concurrent
programs have
been studied for decades~\cite{Volpano_Smith_99,Sabelfeld_Sands_00}. Initial
methods required one to prove that each thread was secure ignorant of the
behaviour of other threads~\cite{Volpano_Smith_99,Sabelfeld_Sands_00,Mantel_Reinhard_07}. Such reasoning is sound but
necessarily imprecise: for instance when reasoning about the top-middle
thread (\autoref{fig:example-mux}) we wouldn't be allowed to assume that
the top-right thread (\autoref{fig:example-demux}) adheres to the locking
protocol that protects~$\buf$.

Following Mantel et al.~\cite{Mantel_SS_11}, more
modern compositional methods have adopted ideas from rely-guarantee
reasoning~\cite{Jones:phd} and concurrent separation logic~\cite{OHearn_04},
to allow more precise
reasoning about each thread under
assumptions it makes about the behaviour of others (e.g.~correct
locking discipline)~\cite{Ernst_Murray_19,Murray_SPR_16,Murray_SE_18}.
However,
the precision of these methods comes at the price of expressiveness:
specifically, their inability to reason about declassification.
%
%
%
By \emph{decoupling} functional
correctness reasoning, \OurSystem achieves both precision and expressiveness.


The \OurSystem logic---\OurSystem's compositional IFC proof method---has
judgements of the form $\HasType{\Alvl}{c}$, where $\Alvl$ is a security level
(e.g.\ $\top$ or $\bot$ in the case of \autoref{fig:example}) representing
level of the attacker
and~$c$
is a fragment of program text (i.e.~a program statement). This judgement holds if the program
fragment~$c$ doesn't leak information to level~$\Alvl$ that $\Alvl$ should not
be allowed to observe. For the code of each thread~$t$, one uses the
rules of \OurSystem's logic to prove that
$\HasType{\Alvl}{c}$ holds, where~$\Alvl$ ranges over all possible security levels.
By doing so one establishes that the concurrent program is secure, under
the assumption that the concurrent program is functionally correct
(i.e.\ each of its annotations~$\ann{A_i}$ hold when the concurrent program
is executed). As mentioned, functional correctness 
can be proved using a range of well-established techniques that integrate
into \OurSystem.

Unlike recent compositional proof methods
(c.f.~\cite{Ernst_Murray_19,Mantel_SS_11,Murray_SPR_16,Murray_SE_18}), the judgement of
\OurSystem has no need to track variable stability information
(i.e.\ which variables won't be modified by other threads), nor
any need for a flow-sensitive
typing context to track the sensitivity of data in shared program
variables, nor does it
track constraints on
the values of program variables.
Instead, this information is provided via the annotations~$\ann{A_i}$.

For example, the annotation $\ann{A_{11}}$ in \autoref{fig:example}
(\autoref{fig:example-demux}, line~4) states that:
(1)~when $\valid$ is 1, if $\inmode$ is 0 then $\buf$ contains the last
input read from channel~$\bot$ and otherwise it contains the last~$\top$ input;
(2) the top-right
thread holds the lock~$\varl$; (3) $\inmode$ and $\outmode$ agree;
and (4)~$\outmode$ is 0 and $\valid$ is 1.
Condition (1) implicitly encodes sensitivity information
about the data in the shared variable~$\buf$; (2) encodes stability
information; while (3) and~(4) are constraints on shared program variables.

To prove that the assignment on line~4 of \autoref{fig:example-demux} is secure,
\OurSystem requires one to show that the sensitivity of the
data contained in~$\buf$ is at most~$\bot$ (the level of $\lbuf$). However
one gets to assume that the annotation at this point~$\ann{A_{11}}$ holds.
In this case, the obligation is discharged straightforwardly from the
annotation. The same is true for other other parts of this concurrent program.
In this way, \OurSystem leans on the functional correctness annotations to
establish security, and utilises compositional functional correctness to
unify reasoning about various security concerns (e.g.\ declassification,
state-dependent sensitivity,~etc.).

\subsection{Proving a Concurrent Program Secure}\label{sec:overview-using}

\begin{figure}
  \includegraphics[width=\columnwidth]{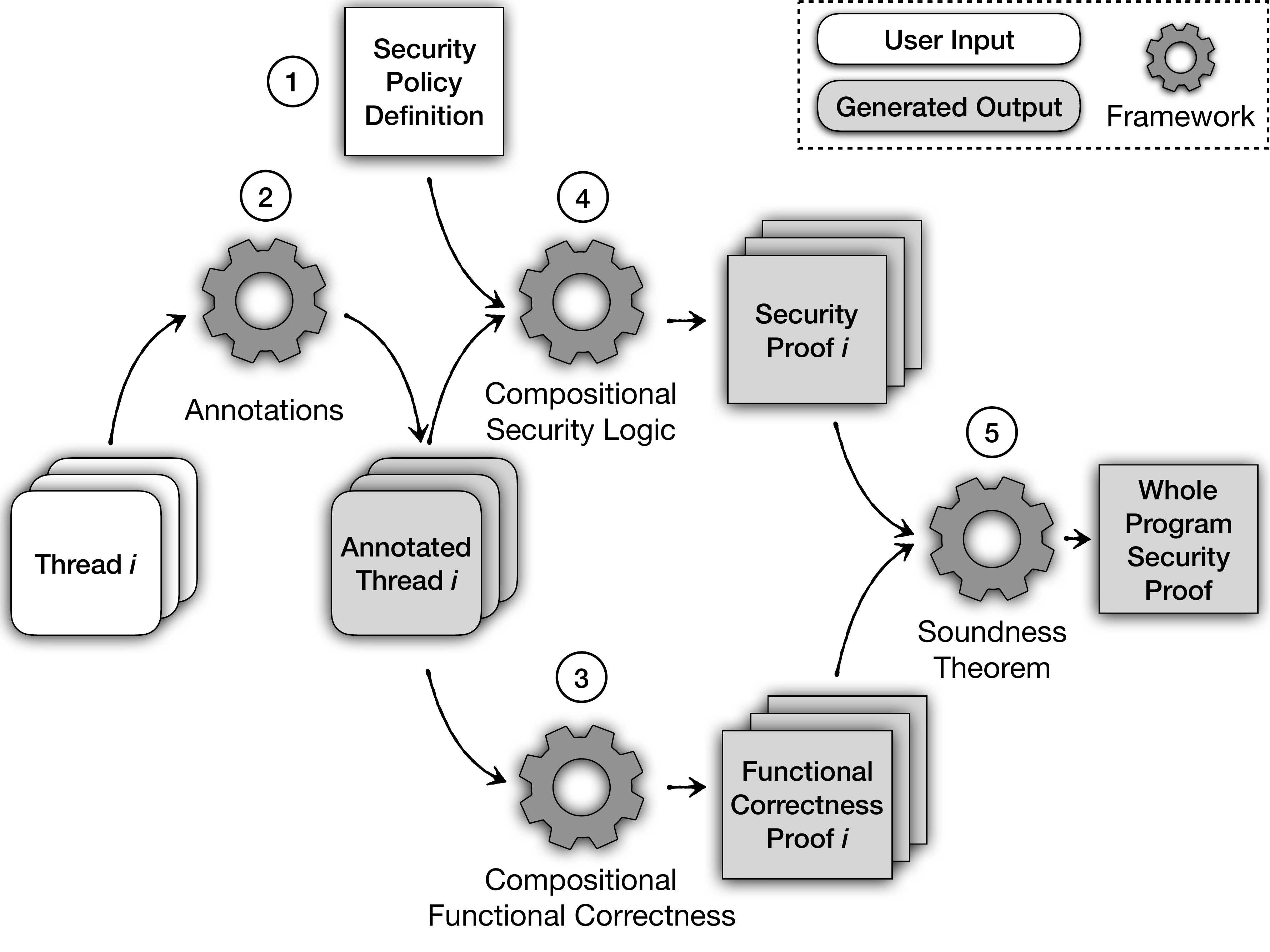}
  \caption{Proving a program secure in \OurSystem.\label{fig:using}}
\end{figure}

\autoref{fig:using} depicts the process of proving a concurrent program
secure using \OurSystem. The circled numbers indicate the main steps and
their ordering.

\subsubsection*{Step \circled{1}: Defining the Security Policy}
The first step is to define the security policy that is to
be enforced. This involves two tasks. The first is to choose
an appropriate
lattice of security levels~\cite{Bell_LaPadula_76} and then to
assign security levels to shared variables
(e.g.\ in the example of \autoref{fig:example}, $\lbuf$ and~$\ldecl$
both have level~$\bot$, while
$\hbuf$ has level~$\top$). A variable's security level is given by the
(user supplied) function~$\L$, which assigns levels to variables.
For variable~$v$, $\L(v)$
defines the maximum
sensitivity of the data that~$v$ is allowed to hold at all times.

In \OurSystem not all shared variables need be assigned a security level,
meaning that $\L$ is allowed to be a \emph{partial} function.
For instance,
in the example of \autoref{fig:example}, $\buf$ has no level assigned
(i.e.\ $\L(\buf)$ is undefined).
The security policy does not restrict the sensitivity of the data that
such \emph{unlabelled} variables are allowed to hold. This is useful for
shared variables like~$\buf$ that form the
interface between two threads and whose sensitivity is governed by
a data-dependent contract~\cite{Murray_SE_18}. In the example,
this allows 
$\buf$ (whenever $\valid$ is 1) to hold $\bot$ data when $\inmode$ and $\outmode$ are both zero,
and $\top$ data when $\inmode$ and $\outmode$ are both nonzero.

The second part of defining the security policy is to specify when and how
declassification is allowed to occur. In order to maximise expressiveness,
\OurSystem supports dynamic,
state-dependent declassification policies. Such policies are encoded via
the (user supplied) predicate~$\D$. For a source security level~$\lsrc$
and destination security level~$\ldst$, the program command~$c$
is allowed to declassify the $\lsrc$-sensitivity value~$v$ to level~$\ldst$
in system state~$\sigma$
precisely when
$\D(\lsrc,\ldst,\sigma,v,c)$ holds. Note that the command~$c$ is either
a declassifying assignment~``\DAssign{$A_i$}{$x$}{$E$}''
(in which case~$\ldst$ is the label~$\L(x)$ assigned to the labelled variable~$x$)
or a
declassifying output~``\DOutput{$A_i$}{$\ldst$}{$E$}''. In either case,
$\lsrc$ is the security level of the expression~$E$ and~$v$ is the result of
evaluating~$E$ in state~$\sigma$.

This style of declassification predicate is able to support
various declassification policies, including delimited release style policies
as in the example of \autoref{fig:example}. We discuss precisely how
delimited release policies are encoded as declassification predicates~$\D$
later in \autoref{sec:dr}. Other declassification policies are encountered
in \autoref{sec:examples}. 

\subsubsection*{Step \circled{2}: Supply Annotations}
Having defined the security policy, the second step to proving a concurrent
program secure using \OurSystem is to supply sufficient functional correctness
annotations~$\ann{A_i}$ for each thread. In the example of \autoref{fig:example}, while their contents is
not shown, these annotations are already present. However in
practice, users of \OurSystem will start with un-annotated programs for which
functional correctness annotations~$\ann{A_i}$
are then supplied to decorate statements~$c$ of each thread, encoding what facts are believed to be true
about the state of the concurrent program whenever statement~$c$ executes.

Note that, because these annotations will be verified later
(in step~\circled{3}), there is no need to trust the process that generates them.
The current Isabelle incarnation of \OurSystem includes a proof-of-concept
strongest-postcondition style annotation inference algorithm,
whose results can then be manually tweaked by the user as necessary.
Users are also free to employ external,
automatic program analysis tools to infer functional correctness annotations,
or to supply annotations manually,
without fear of compromising the foundational guarantees of \OurSystem.

\subsubsection*{Step \circled{3}: Verifying Functional Correctness}
\label{sec:concurrent-functional-correctness}
Having obtained the functional correctness annotations~$\ann{A_i}$,
the next step is
to \emph{prove} their validity.
This means proving that the concurrent program is functionally correct,
for which there exist numerous compositional techniques~\cite{Owicki_Gries_76,Jones:phd}.

\OurSystem incorporates two standard techniques in the Isabelle/HOL formalisation: the Owicki-Gries
method~\cite{Owicki_Gries_76} and Rely-Guarantee reasoning~\cite{Jones:phd}.
\OurSystem's Owicki-Gries implementation is borrowed from the seminal work
of Prensa~Nieto~\cite{Nieto_Esparza_00,Prensa:phd}. Using it to verify
(correct) functional correctness annotations 
requires little effort for experienced Isabelle users, by
guiding Isabelle's proof automation tactics.
Like \OurSystem's Owicki-Gries method, its Rely-Guarantee
implementation is for verifying functional correctness annotations only,
and ignores security (c.f.~\cite{Mantel_SS_11, Murray_SPR_16}). It
requires the user to supply \emph{rely} and \emph{guarantee} conditions for
each thread. Such conditions can be defined straightforwardly from the locking
protocol of a concurrent program and in principle could be inferred; however,
we leave that inference for future work.

If one wishes to forego the foundational assurance of Isabelle/HOL, one can
also employ external verification tools to prove annotation validity. By
doing so
one may elide \emph{non-essential} annotations, which are those
on statements other than
output statements, declassifications, assignments to labelled variables, and branches
on unlabelled data.
Our formalisation includes a proof-of-concept of this approach, in which the
concurrent C program verifier~VCC~\cite{DBLP:conf/tphol/CohenDHLMSST09} is
employed on a C translation of the example in \autoref{fig:example} to prove
the functional correctness of its essential annotations, sufficient to
guarantee its security.

%

\subsubsection*{Step \circled{4}: Verifying Security}
With functional correctness proved, the user is then free to use the
functional correctness annotations to compositionally prove the security
of the concurrent program. To do this, the user applies the rules of the
\OurSystem logic to each of the program's threads.
\OurSystem exploits the functional correctness assertions to provide a simple
logic resembling a flow-insensitive type system. Each statement is verified
independently of its context. The logic is compositional and allows each
thread to be verified in isolation. Rules for output statements require that
the functional correctness annotations imply that the expression always
evaluates to the same result given states that are not distinguishable to the
attacker. Similarly, the rules for declassification statements require that
the annotations are sufficient to imply that the declassification predicate
holds.
We defer a full presentation of the logic to \autoref{sec:logic}.

\subsubsection*{Step \circled{5}: Whole Program Security Proof}
With both functional correctness and security proved of each thread,
the soundness
theorem of the \OurSystem logic can then be applied to derive a theorem stating
that the whole concurrent program is secure. This theorem is stated formally
in \autoref{sec:soundness}. However, intuitively it says that the whole
concurrent program is secure if, for each thread~$t$, $t$'s functional
correctness annotations are all valid (i.e.\ each holds whenever the
corresponding statement of~$t$ is executed in the concurrent program)---step~\circled{3}---and $t$ is judged secure by the rules of the \OurSystem logic---step~\circled{4}.

\section{Security Definition}
\label{sec:security-property}

\OurSystem proves an information flow security property designed
to capture a range of different security policies. To maximise
generality, the security property is phrased in a knowledge-based (or
\emph{epistemic}) style, which as others have
argued~\cite{Broberg_vDS_15,Askarov_Sabelfeld_07,Askarov_Chong_12,Broberg_Sands_09}
is preferable to
traditional two-run formulations.
Before introducing the security property and motivating the threat model that
it formally encodes, we first explain the semantic
model of concurrent program execution in which the property is~defined.

%

Along the way, we highlight the assumptions encoded in that semantic model
and in the formal security property. Following Murray and
van~Oorschot~\cite{Murray_vanOorschot_18}, we distinguish
\emph{adversary expectations}, which are
assumptions about the attacker (e.g.\ their observational powers);
from \emph{domain hypotheses}, which
are assumptions about the environment (e.g.\ the scheduler)
in which the concurrent program
executes.

\subsection{Semantic Model}\label{sec:model}

Concurrent programs comprise a finite collection of~$n$ threads, each
of which is identified by a natural number: $0,\ldots,n-1$. Threads
synchronise by acquiring and releasing locks and communicate by modifying
shared memory. Additionally, threads may communicate with
the environment outside the concurrent program by inputting and outputting
values from/to IO \emph{channels}. Without loss of generality,
there is one channel for each security
level (drawn from the user-supplied lattice of security levels).

\subsubsection*{Global States $\sigma$}
Formally, the \emph{global states}~$\sigma$ of the concurrent program
are tuples~$\gstate{\sigma}$. The global state
contains all resources that are shared between threads. We consider each
in turn.

\subsubsection*{Channels and the Environment~$\env[\sigma]$}
$\env[\sigma]$ captures the state of the external
environment (i.e.\ the IO channels). For a security level~$\lvl$,
$\env[\sigma](\lvl)$ is the (infinite) stream of values yet to be consumed
from the channel~$\lvl$ in state~$\sigma$.
\begin{domainhypothesis}
  In this model of channels, 
  reading from a channel never blocks and always returns the next value
  to be consumed from the infinite stream. This effectively assumes that all
  channel inputs are faithfully buffered and never dropped by the environment.
  Blocking can be simulated
  by repeatedly polling a channel.
\end{domainhypothesis}
\vspace{-2ex}
\subsubsection*{Shared Memory~$\mem[\sigma]$}
$\mem[\sigma]$ is simply a total mapping form variable names (excluding locks)
to
their corresponding values: $\mem[\sigma](v)$ denotes the value of variable~$v$
in state~$\sigma$.

\subsubsection*{Locks~$\locks[\sigma]$}
$\locks[\sigma]$ captures the lock state and is a partial
function from lock names to thread ids (natural numbers in the range $0\ldots{n-1}$): for a
lock~$\varl$, $\locks[\sigma](\varl)$ is defined iff lock~$\varl$ is currently
held in state~$\sigma$, in which case its value is the id of the thread that
holds the lock.

\subsubsection*{Events~$e$ and Traces~$\trace[\sigma]$}
For the sake of expressiveness, we store in the global state~$\sigma$ the entire
history of events~$\trace[\sigma]$ that has been performed by the concurrent
program up to this point. Each such history is called a \emph{trace}, and
is simply a finite list of events~$e$.  Events~$e$ comprise:
\emph{input} events $\ievent{\lvl}{v}$ which record that value~$v$ was input from
the channel~$\lvl$; \emph{output} events $\oevent{\lvl}{v}{E}$ which record
that value~$v$, the result of evaluating expression~$E$, was output
on channel~$\lvl$; and \emph{declassification} events $\devent{\lvl}{v}{E}$
which record that the value~$v$, the result of evaluating expression~$E$,
was declassified to level~$\lvl$. Expression~$E$ is included to help 
specify the security property (see e.g.~\autoref{defn:event-secure}).

Ordinary (non-declassifying) output 
and input commands produce output and input events respectively.
Declassifying assignments and
declassifying outputs produce 
declassification events.
As with much prior work on declassification properties~\cite{Cecchetti_MO_17},
declassification actions produce distinguished declassification events that
make them directly visible to the security property. 

\subsubsection*{The Schedule~$\sched$}\label{sec:sched}

The \emph{schedule}~$\sched$ is an infinite list (stream) of
thread ids~$i$. Scheduling occurs by removing the first item~$i$ from the
stream and
then executing the thread~$i$ for one step of execution. (Longer execution
slices can of course be simulated by repeating~$i$ in the schedule.)
This process is repeated \emph{ad infinitum}. If thread~$i$ is stuck
(e.g.\ because it is waiting on a lock or has terminated) then the system
idles (i.e.\ does nothing) for an execution step, to mitigate scheduling leaks (e.g.~as implemented
in seL4~\cite{Murray_MBGBSLGK_13}).
\begin{domainhypothesis}
  \OurSystem assumes deterministic, sequentially-consistent, instruction-based scheduling~\cite{Stefan_BYLTRM_13}
  (IBS) of
  threads against a fixed, public schedule. 
\end{domainhypothesis}

\vspace{-3ex}
\subsubsection*{Global Configurations and Concurrent Execution~$\gstep{\cdot}{\cdot}$}

A \emph{global configuration} combines the shared global state~$\sigma$ with
the schedule~$\sched$ and the local state~$\ls[i]$ (the thread id and code)
of each of the~$n$ threads.
Thus a global configuration is a
tuple: $(\ls[0],\ldots,\ls[n-1],\sigma,\sched)$. 

Concurrent execution, and the aforementioned scheduling model,
\ifExtended
is
formally defined by the rules of \autoref{fig:global-exec} (relegated to the
appendix for brevity).
\fi
These rules
define a single-step relation~$\gstep{\cdot}{\cdot}$ on global configurations.
Zero- and multi-step execution is captured in the usual way by the reflexive,
transitive closure of this relation, written~$\gstepstar{\cdot}{\cdot}$.

\subsection{System Security Property and Threat Model}\label{sec:threat-model}

We now define \OurSystem's formal security property,
formalising the threat model and adversary
expectations. 

\subsubsection*{Attacker Observations}
\vspace{-4ex}
\noindent\begin{adversaryexpectation}
Our security property considers a passive attacker
observing the execution of the concurrent program.
We assume that the
attacker is able to observe outputs on certain channels and associated
declassification events.
Specifically, the attacker is associated with
a security level~$\Alvl$. Outputs on all channels~$\lvl \leq \Alvl$ the
attacker is assumed to be able to observe. Likewise all declassifications to
levels~$lvl \leq \Alvl$.
\end{adversaryexpectation}
\vspace{-2ex}
\begin{adversaryexpectation}
The attacker has no other means to
interact with the concurrent program, e.g.\ by modifying its code.
We additionally assume that the attacker does not have access to timing
information.
\end{adversaryexpectation}
The attacker's observational powers are formalised by defining a series of
indistinguishability relations as follows.

\begin{definition}[Event Visibility]
We say that an input event~$\ievent{\lvl}{v}$ (respectively output event~$\oevent{\lvl}{v}{E}$ or declassification event~$\devent{\lvl}{v}{E}$) is
\emph{visible} to the attacker at level~$\Alvl$ iff $\lvl \leq \Alvl$.
Letting~$e$ be the event, in this case we write $\visible{\Alvl}(e)$.
\end{definition}

Trace indistinguishability is then defined straightforwardly, noting that
we write $\filter{P}{\trace}$ to denote filtering from trace~$\trace$ all events
that do not satisfy the predicate~$P$.

\begin{definition}[Trace Indistinguishability]
  We say that two traces~$\trace$ and~$\trace'$ are indistinguishable to the
  attacker at level~$\Alvl$, when $\filter{\visible{\Alvl}}{\trace} = \filter{\visible{\Alvl}}{\trace'}$.

  In this case, we write $\trequiv{\Alvl}{\trace}{\trace'}$.
\end{definition}

\subsubsection*{Attacker Knowledge of Initial Global State}
Besides defining what the attacker is assumed to observe (via the
indistinguishability relation on traces), we also need to define what
knowledge the attacker is assumed to have about the initial global state~$\sigmainit$ of the
system.
\begin{adversaryexpectation}
The attacker
is assumed to know the contents that will be input from channels at
levels~$\lvl \leq \Alvl$
and the initial values of all labelled variables~$v$ for which
$\L(v) \leq \Alvl$.
\end{adversaryexpectation}
This assumption is captured via an indistinguishability relation on global
states~$\sigma$. This relation is defined by first defining indistinguishability
relations on each of $\sigma$'s components.

\begin{definition}[Environment Indistinguishability]
  We say that two environments~$\env$ and~$\env'$ are indistinguishable
  to the attacker at level~$\Alvl$ when
  all channels visible to the attacker have identical
  streams, i.e.\ iff
  $$\forall \lvl \leq \Alvl.\ \env(\lvl) = \env'(\lvl).$$ In this case we write~$\envequiv{\Alvl}{\env}{\env'}$.
\end{definition}

\begin{definition}[Memory Indistinguishability]
  We say that two memories~$\mem$ and~$\mem'$ are indistinguishable to
  the attacker at level~$\Alvl$ when they agree on the values of
  all labelled variables~$v$ visible to the attacker,
  i.e.\ iff
  $$\forall v.\  \L(v) \leq \Alvl \implies \mem(v) = \mem'(v),$$
  where $\L(v) \leq \Alvl$ implies $\L(v)$ is defined.

  \noindent In this case, we write $\memequiv{\Alvl}{\mem}{\mem'}$.
\end{definition}

We can now define when two (initial) global states are indistinguishable
to the attacker.

\begin{definition}[Global State Indistinguishability]
  We say that two global states~$\sigma$ and~$\sigma'$ are indistinguishable
  to the attacker at level~$\Alvl$ iff
  \[
  \begin{array}{l}
  \envequiv{\Alvl}{\env[\sigma]}{\env[\sigma']} \mathrel{\land}
  \memequiv{\Alvl}{\mem[\sigma]}{\mem[\sigma']} \mathrel{\land} \\
  \locks[\sigma] = \locks[\sigma'] \mathrel{\land}
  \trequiv{\Alvl}{\trace[\sigma]}{\trace[\sigma']}
  \end{array}
  \]
  In this case we write $\gequiv{\Alvl}{\sigma}{\sigma'}$.
\end{definition}
\begin{domainhypothesisnobreak}
  Under this definition, the attacker knows the entire initial
  lock state. Thus we assume that the initial lock state encodes no secret information.
\end{domainhypothesisnobreak}

\vspace{-2ex}
\subsubsection*{Attacker Knowledge from Observations}
Given the attacker's knowledge about the initial state~$\sigmainit$ and some
observation arising from some trace~$\trace$ being performed,
we assume that the attacker will then attempt to refine
their knowledge about~$\sigmainit$.
\begin{adversaryexpectation}
The attacker is assumed to know the
schedule~$\sched$ and the initial local state~$\ls[i]$ (i.e.\ the code and thread id~$i$) of each thread.
\end{adversaryexpectation}
Given that information, of all the
possible initial states from which~$\sigmainit$ might have been drawn, perhaps
only a subset can give rise to the observation of~$\trace$. We assume the
attacker will perform this kind of knowledge
inference, which we formalise following
the epistemic style~\cite{Askarov_Sabelfeld_07}.

To define the attacker's knowledge, we define the attacker's uncertainty about the initial
state~$\sigmainit$ (i.e.\ the attacker's belief about the set of all
initial states from which~$\sigmainit$ might have been drawn) given
the initial schedule~$\sched$ and local thread states~$\ls[0],\ldots,\ls[n-1]$, and the trace~$\trace$ that the attacker has observed.
Writing simply $\ls$ to abbreviate the list~$\ls[0],\ldots,\ls[n-1]$,
we denote this $\uncertainty{\Alvl}{\ls}{\sigmainit}{\sched}{\trace}$ and define
it as follows.

\begin{definition}[Attacker Uncertainty]
  A global state~$\sigma$ belongs to the set
  $\uncertainty{\Alvl}{\ls}{\sigmainit}{\sched}{\trace}$ iff it and
  $\sigmainit$ are 
  indistinguishable, given the attacker's knowledge
  about the initial state, and if $\sigma$ can give rise to a trace
  $\trace[\sigma']$
  that is indistinguishable from~$\trace$.
  Formally, $\uncertainty{\Alvl}{\ls}{\sigmainit}{\sched}{\trace}$ is the set of~$\sigma$~where
  \[
  \begin{array}{l}\gequiv{\Alvl}{\sigma}{\sigmainit} \mathrel{\land} \\
        \exists \ls'\ \sigma'\  \sched'.\ \gstepstar{(\ls,\sigma, \sched)}{(\ls',\sigma',\sched')} \mathrel{\land} \\ \qquad \qquad \qquad \quad  \trequiv{\Alvl}{\trace}{\trace[\sigma']} \end{array}
  \]
\end{definition}

\subsubsection*{The Security Property}

Finally, we can define the security property. This requires roughly that
the attacker's uncertainty can decrease (i.e.\ they can refine their
knowledge) only when declassification events occur, and that all such
events must respect the declassification policy encoded by~$\D$.
In other words, the guarantee provided by \OurSystem under the threat
model formalised herein is that:
\begin{securityguarantee}
The attacker is never able to learn
any new information above what they knew initially, \emph{except} from
declassification events but those must \emph{always} respect
the user-supplied declassification policy.
\end{securityguarantee}

This guarantee is formalised by defining a \emph{gradual release}-style
security property~\cite{Askarov_Sabelfeld_07}.
We first define when the occurrence of an event~$e$ is secure.

\begin{definition}[Event Occurrence Security]\label{defn:event-secure}
  Consider an execution beginning in some initial configuration~$(\ls[],\sigma,\sched)$ that has executed to the intermediate configuration~$(\ls[]',\sigma',\sched')$ from which the event~$e$ occurs. This occurrence is secure against
  the attacker at level~$\Alvl$, written~$\eventsecure{\Alvl}{\ls[]}{\sigma}{\sched}{\ls[]'}{\sigma'}{\sched'}{e}$, iff
  \begin{itemize}
  \item When~$e$ is a declassification event~$\devent{\ldst}{v}{E}$
    visible to the attacker (i.e.\ $\ldst \leq \Alvl$), then
    $\D(\L(E),\ldst,\sigma',v,c)$ must hold, where~$c$ is the current
    program command whose execution produced~$e$ (i.e.\ the head program
    command of the currently executing thread in $(\ls[]',\sigma',\sched')$).
    Here, $\L(E)$ is defined when $\L(v)$ is defined for all variables~$v$
    mentioned in~$E$ and in that case is the least upper bound of
    all such~$\L(v)$, and $\D(\L(E),\ldst,\sigma',v,c)$ is false
    when $\L(E)$ is not~defined.
  \item Otherwise, if~$e$ is not a declassification
    event~$\devent{\ldst}{v}{E}$ that is visible to the attacker,
    then the attacker's uncertainty cannot decrease by observing it,
    i.e. we require that
    \[
    \begin{array}{l}
      \uncertainty{\Alvl}{\ls[]}{\sigma}{\sched}{\trace[\sigma']} \subseteq \\
      \;\; \uncertainty{\Alvl}{\ls[]}{\sigma}{\sched}{\trace[\sigma'] \cdot e}
    \end{array}
    \]
  \end{itemize}
\end{definition}

\begin{definition}[System Security]\label{defn:sys-secure}
  The concurrent program with initial local thread states~$\ls[] = (\ls[0],\ldots,\ls[n-1])$ is secure against an attacker at level~$\Alvl$,
  written $\syssecure{\Alvl}{\ls[]}$,
  iff, under all schedules~$\sched$,
  event occurrence security always holds during its execution from any
  initial starting state~$\sigma$. Formally, we require that
  \[
  \begin{array}{l}
    \forall \sched\ \sigma\ \ls[]'\ \sigma'\ \sched'\ \ls[]''\ \sigma'' \sched''\ e. \\
    \;\;\;\gstepstar{(\ls[], \sigma,\sched)}{(\ls[]', \sigma', \sched')} \mathrel{\land} \\
    \;\;\;\gstep{(\ls[]', \sigma', \sched')}{(\ls[]'', \sigma'', \sched'')} \mathrel{\land} \\
    \;\;\;\trace[\sigma''] = \trace[\sigma'] \cdot e \mathrel{\implies} \\
    \;\;\;\;\;\;\eventsecure{\Alvl}{\ls[]}{\sigma}{\sched}{\ls[]'}{\sigma'}{\sched'}{e}
  \end{array}
  \]
\end{definition}

\subsection{Discussion}\label{sec:security-prop-discussion}
As with other gradual release-style properties,
ours does not directly constrain \emph{what} information the attacker might
learn when a declassification event occurs, but merely that those are the
only events that can increase the attacker's knowledge. This means that,
of the four semantic principles of declassification
identified by Sabelfeld and Sands~\cite{Sabelfeld_Sands_09}, our definition
satisfies all but \emph{non-occlusion}: ``the presence of
declassifications cannot mask other covert information leaks''~\cite{Sabelfeld_Sands_09}.
Consider the following single-threaded program.

\medskip
\newcommand{\birthYear}{\mathit{birthYear}}
\newcommand{\birthDay}{\mathit{birthDay}}
\newcommand{\birthMonth}{\mathit{birthMonth}}
\noindent\begin{examplecode}
  \ITENoEndIf{\nextann}{$\birthYear > 2000$}{\DOutput{\nextann}{$\bot$}{$\birthDay$}}{\DOutput{\nextann}{$\bot$}{$\birthMonth$}}
\end{examplecode}
\medskip\\
Suppose the intent is to permit the unconditional
release of a person's day and month of
birth, but not their birth year. A naive encoding in the declassification
policy~$\D$ that checks whether the value being declassified is indeed
either the value of~$\birthDay$ or~$\birthMonth$ would judge the above program
as secure, when in fact it also leaks information about the
$\birthYear$.

Note also, since declassification events are directly visible to our
security property, that programs
that incorrectly
declassify information but then never output it on a public channel can be
judged by our security condition as insecure.

Finally, and crucially, note that our security condition allows for both
\emph{extensional}
declassification policies, i.e. those that refer only to inputs and
outputs of the program, as well as \emph{intensional} policies that also refer
to the program state. \autoref{sec:examples} demonstrates both kinds of policies. We now consider one class of extensional policies: delimited release.  

\subsection{Encoding Delimited Release Policies}\label{sec:dr}

The occlusion example demonstrates that programs that branch on secrets
that are not allowed to be released
and then perform declassifications under that secret context are likely to
leak more information than that contained in the declassification events
themselves, via implicit flows.

However,
in the absence of such branching, our security condition
can in fact place bounds on what information is released.
Specifically, we show that it can soundly encode delimited release~\cite{Sabelfeld_Myers_03} policies as
declassification predicates~$\D$ for programs that do not branch on
secrets that are not allowed to be declassified to the attacker.

We define an extensional delimited release-style security condition and show how to
instantiate the declassification predicates~$\D$ so that when
system security (\autoref{defn:sys-secure}) holds, then so does the
delimited release condition.

\subsubsection{Formalising Delimited Release}

Delimited release~\cite{Sabelfeld_Myers_03} weakens traditional
noninterference~\cite{Goguen_Meseguer_82} by permitting certain
secret information to be released to the attacker. Which secret
information is allowed to be released is defined in terms of a set
of \emph{escape hatches}: expressions that denote values allowed
to be released.

Delimited release then strengthens the indistinguishability relation on the
initial state to require that any two states related under this relation
also agree on the values of the escape hatch expressions. One way to
understand delimited release as a weakening of noninterference is to observe
that, in changing the relation in this way, it is effectively
encoding the assumption that the attacker might \emph{already know} the
secret information denoted by the escape hatch expressions.

To keep our formulation brief, we assume that the initial memory contains
no secrets. Thus all secrets are contained only in the
input streams (channels).
Then escape hatches denote values that are allowed to
be released as functions on lists~$\vs$ of inputs (to be) consumed from a channel.

A \emph{delimited release} policy~$\E$ is a function that 
given source and destination security levels~$\lsrc$ and~$\ldst$ returns
a set of escape hatches denoting the information that is allowed to be
declassified from level~$\lsrc$ to level~$\ldst$.

For example, to specify that the program is always allowed to declassify
to~$\bot$
the average of the last five inputs read from the~$\top$ channel, one
could define
  $\E(\top,\bot) =
  \{\lambda \vs.\ \IfThenElse{\length(vs) \geq 5}{\avg(\take(5,\rev(\vs)))}{0}\}$,
where $\avg(\xs)$ calculates the
average of a list of values~$\xs$,
$\take(n,\xs)$  returns a new list
containing the first~$n$
values from the list~$\xs$, and $\rev(xs)$ is the list reversal
function.

To define delimited release, we need to define when two initial states~$\sigma$
agree under the escape hatches~$\E$. Since escape hatches apply only to the
streams contained in the environment~$\env[\sigma]$, we define when two
such environments agree under~$\E$. As earlier,
this agreement is defined relative to an attacker observing
at level~$\Alvl$, and requires that all escape hatches that yield values
that the attacker is allowed to observe always evaluate identically under
both environments.

\begin{definition}[Environment Agreement under~$\E$]
  Two environments~$\env[]$ and~$\env[]'$ agree under
  the delimited release policy~$\E$ for an attacker at level~$\Alvl$,
  written \mbox{$\envdrequiv{\Alvl}{\E}{\env[]}{\env[]'}$,} iff, for all
  levels~$\lsrc$ and all
  levels~$\ldst \leq \Alvl$, and escape hatches~$h\in\E(\lsrc,\ldst)$,
  $h$ applied to any finite prefix of~$\env[](\lsrc)$ yields the same
  value as when applied to an equal length prefix of~$\env[]'(\lsrc)$.
\end{definition}

We then define when two initial states~$\sigma$ and~$\sigma'$
agree for a delimited release policy~$\E$.
The following
definition is a slight simplification of the one in our Isabelle formalisation
(see \ifExtended\autoref{defn:dr-equiv-full} in the appendix\fi),
which is more general because it considers arbitrary pairs of states
in which some
trace of events might have already been performed.  

\begin{definition}[State Agreement under $\E$]\label{defn:dr-equiv}
  States~$\sigma$ and~$\sigma'$ agree under
  the delimited release policy~$\E$ for an attacker at level~$\Alvl$,
  written $\gdrequiv{\Alvl}{\E}{\sigma}{\sigma'}$, iff
  (1)~$\gequiv{\Alvl}{\sigma}{\sigma'}$,
  (2)~their memories
  agree on all variables, and (3)~$\envdrequiv{\Alvl}{\E}{\env[\sigma]}{\env[\sigma']}$.
\end{definition}
\noindent Here, condition~(2) encodes the simplifying assumption that the initial
memories contain no secrets.

Delimited release is then defined extensionally in the style of
traditional two-run noninterference property. 
\begin{definition}[Delimited Release]\label{defn:dr-secure}
  The concurrent program with initial local thread states~$\ls[] = (\ls[0],\ldots,\ls[n-1])$ satisfies \emph{delimited release} against an attacker at
  level~$\Alvl$,
  written $\drsecure{\Alvl}{\ls[]}$,
  iff, 
  \[
  \begin{array}{l}
    \forall \sched\ \sigma\ \sigma'\ y.\ 
          \gdrequiv{\Alvl}{\E}{\sigma}{\sigma'} \mathrel{\land} 
          \gstepstar{(\ls[], \sigma,\sched)}{y} \mathrel{\implies} \\
    \;\;\;\;\;\;\;\;\;\;\;\;\;\;\;\;\;\;\;\;\;\;\;\;(\exists y'.\ \gstepstar{(\ls[], \sigma', \sched)}{y'} \mathrel{\land} 
    \trequiv{\Alvl}{\trace[y]}{\trace[y']})
  \end{array}
  \]
  where for a global configuration $y = (\ls[y],\sigma_y,\sched_y)$ we
  write $\trace[y]$ to abbreviate $\trace[\sigma_y]$, the trace executed so far.
\end{definition}

\subsubsection{Encoding Delimited Release in $\D$}

We now encode delimited release policies~$\E$ via
\OurSystem's declassification predicates~$\D(\lsrc,\ldst,\sigma,v,c)$
which, recall, judge whether command~$c$
declassifying value~$v$ from level~$\lsrc$
to level~$\ldst$ in state~$\sigma$ is permitted.
Recall that~$c$
is either
a declassifying assignment~``\DAssign{$A_i$}{$x$}{$E$}''
(in which case~$\ldst$ is the label~$\L(x)$ assigned to the labelled variable~$x$)
or a
declassifying output~``\DOutput{$A_i$}{$\ldst$}{$E$}''. In either case,
$\lsrc$ is the security level of the expression~$E$ and~$v$ is the result of
evaluating~$E$ in state~$\sigma$.

To encode delimited release, we need to have
$\D(\lsrc,\ldst,\sigma,v,c)$ decide whether there is an escape hatch
$h \in \E(\lsrc,\ldst)$
that permits the declassification. Consider some $h \in  \E(\lsrc,\ldst)$.
What does it mean for $h$ to permit the declassification?
Perhaps surprisingly, it is \emph{not enough} to
check whether $h$ evaluates to the value~$v$ being declassified in~$\sigma$.
Suppose $h$ permits declassifying the average of the last five inputs
from channel~$\top$ and suppose in~$\sigma$ that this average is 42.
An insecure program might declassify some \emph{other} secret whose value just
happens to be 42 in~$\sigma$, but that declassification would be
unlikely to satisfy
delimited release if the two secrets are independent.

Instead, to soundly encode delimited release, one needs to check whether
the expression~$E$ being declassified is equal to the escape hatch
\emph{in general}. 

To do this, we have $\D$ check that in all states in which this
declassification~$c$ might be performed, the escape hatch~$h$ evaluates
to the value of~$E$ in that state. We can overapproximate the set of all
states in which~$c$ might execute by using
its annotation~$\ann{A_i}$:
all such states must satisfy the annotation assuming
the program is functionally correct
(which \OurSystem will prove).
Thus we have $\D$ check that
in all such states that satisfy the annotation, the escape hatch~$h$ evaluates
to the expression~$E$.


\begin{definition}[Delimited Release Encoding]\label{defn:dr-encoding}
  The encoding of policy~$\E$
  we denote~$\D_\E$.
  $\D_\E(\lsrc,\ldst,\sigma,v,c)$ holds always when~$c$ is not a
  declassification command. Otherwise, let~$A$ be $c$'s annotation
  and~$E$ be the expression that~$c$ declassifies.
  Then $\D_\E(\lsrc,\ldst,\sigma,v,c)$ holds iff there exists some
  $h \in \E(\lsrc,\ldst)$ such that for all states~$\sigma'$
  that satisfy the annotation~$A$, $E$ evaluates in~$\sigma'$ to the same
  value that~$h$ evaluates to when applied to the $\lsrc$ inputs consumed
  so far in~$\sigma'$.
\end{definition}

Recall this encoding is sound
only for programs that do not branch on secrets that the
policy~$\E$ forbids from releasing. We define this condition semantically as
a two-run property, relegating it to \ifExtended\autoref{defn:no-secret-branching} in the appendix \fi since its meaning is intuitively clear.
We say that a program satisfying this condition
is \emph{free of $\E$-secret branching}.

The example of \autoref{sec:security-prop-discussion} that leaks $\birthYear$
via occlusion is not free of $\E$-secret branching.
On the other hand, the program in
\autoref{fig:example} is free of $\E$-secret branching
for the following~$\E$
that defines its delimited release policy,
since the only $\top$-value ever branched on (in \autoref{fig:example-demux}, line~8) is the result of the signature check~$\CHECKSIG$.

\begin{definition}[Delimited Release policy for \autoref{fig:example}]
  \label{defn:example-dr}
  Allow to be declassified to~$\bot$
  the results of the signature check
  $\CHECKSIG$ always, plus any $\top$-input~$v$ when $\CHECKSIG(v) = 0$.
  \[
\begin{array}{l}
  \E(\top,\bot) = \\
  \;\; \{\lambda \vs.\ \IfThenElse{\length(\vs) \not= 0}{\CHECKSIG(\last(\vs))}{0}\} \union \\
  \;\; \{\lambda \vs.\ \IfThenElse{\length(\vs) \not= 0 \mathrel{\land} \CHECKSIG(\last(\vs)) = 0}{\\\;\;\;\;\last(\vs)}{0}\}
\end{array}
\]
\end{definition}

Indeed, \OurSystem can be used to prove that \autoref{fig:example} satisfies this
delimited release policy by showing that it satisfies \OurSystem's
system security (\autoref{defn:sys-secure}), under the following theorem
that formally justifies why \OurSystem can encode delimited release policies.

\begin{theorem}[Delimited Release Embedding]\label{thm:dr-embedding}
  Let~$\Alvl$ be an arbitrary security level and~$\ls$ be the initial local
  thread states (i.e.\ thread ids and the code) of a concurrent program
  that~(1) satisfies
  $\syssecure{\Alvl}{\ls}$ with $\D$ defined according to \autoref{defn:dr-encoding}, (2)~is free of $\E$-secret branching,
  and~(3) satisfies all of its functional correctness annotations.
  Then, the program is delimited release secure, i.e.\ $\drsecure{\Alvl}{\ls}$.
\end{theorem}

\noindent Thus \OurSystem can soundly encode purely extensional
security properties like \autoref{defn:dr-secure}.
%
The extensional form of the policy 
for the \autoref{fig:example} example is
straightforward and relegated to the \ifExtended appendix 
(\autoref{defn:example-dr-extensional})
\fi.

\section{Annotated Programs in \OurSystem}\label{sec:lang}

\OurSystem reasons about the security of concurrent programs, each of whose
threads is programmed in the language
whose grammar is given in \autoref{fig:lang}. 

\begin{figure}
  \begin{tabular}{r@{\;}lp{4cm}}
    $c$ ::= & \Assign{$A$}{$x$}{$E$} & \small(assignment)\\
    $|$ & \DAssign{$A$}{$x$}{$E$} & \small(declassifying assignment)\\
    $|$ & \Output{$A$}{$\lvl$}{$E$} & \small(output to channel~$\lvl$)\\
    $|$ & \DOutput{$A$}{$\lvl$}{$E$} & \small(declassifying output)\\
    $|$ & \Input{$A$}{$x$}{$\lvl$} & \small(input from channel~$\lvl$)\\
    $|$ & \ITEg{$A$}{$E$}{$c$}{$c$} & \small(conditional)\\
    $|$ & \Whileg{$A$}{$E$}{$A$}{$c$} & \small(loop with invariant)\\
    $|$ & \Acquire{$A$}{$\varl$} & \small(lock acquisition)\\
    $|$ & \Release{$A$}{$\varl$} & \small(lock release) \\
    $|$ & \Seqg{$c$}{$c$} & \small(sequencing) \\
    $|$ & \Skip & \small(terminated thread)
  \end{tabular}
  \caption{Syntax of \OurSystem threads.\label{fig:lang}}
\end{figure}

Most of these commands are straightforward and
appear in \autoref{fig:example}.
Loops ``\Whileg{$A$}{$E$}{$I$}{$c$}'' carry a second
\emph{invariant} annotation (here~``$\ann{I}$'') that specifies the loop
invariant, which is key for proving their functional correctness~\cite{Floyd_67}. The ``\Skip'' command halts the execution of the thread, and
is an
internal form used only to define the semantics of the language.
The \emph{no-op} command
``\Nop{$A$}'' is syntactic sugar for: ``\Assign{$A$}{$x$}{$x$}'', while
``\ITNoElseg{$A$}{$E$}{$c$}'' is  sugar for
``\ITEg{$A$}{$E$}{$c$}{\Nop{$A$}}''.

The semantics for this sequential language is
\ifExtended
given in \autoref{fig:semantics},
and
is relegated to the appendix
\fi
since it is straightforward.
This semantics is defined as a small step relation on
\emph{local configurations}~$(\ls[i],\sigma)$
where~$\ls[i] = (i,c)$ is the local state (thread id~$i$ and code~$c$)
for a thread
and~$\sigma = \gstate{\sigma}$
is the global state shared with all other threads.
$\eval{\mem[\sigma]}{E}$ is atomic evaluation of expression~$E$ in memory~$\mem[\sigma]$. 
Notice that
the semantics doesn't make use of the annotations~$\ann{A}$: they are
merely decorations used to decouple functional correctness.

\section{The \OurSystem Logic}\label{sec:logic}

The \OurSystem logic defines a compositional method to prove when a
concurrent program satisfies system security (\autoref{defn:sys-secure}),
\OurSystem's security condition. Specifically, it defines a set of rules
for reasoning over the program text of each thread of the concurrent
program. A soundness theorem (\autoref{thm:soundness}) guarantees
that programs that are functionally correct and whose threads
are proved secure using the
\OurSystem logic satisfy system security.

The rules of the \OurSystem logic appear in \autoref{fig:logic}.
They define a judgement resembling that for a
flow-insensitive security type system that has the form:
$\HasType{\Alvl}{c}$
where~$\Alvl$ is the attacker level and~$c$ is an annotated thread command
(see \autoref{fig:lang}).

\begin{figure*}
  \begin{mathpar}
    \inferrule{\HasType{\Alvl}{c_1} \\ \HasType{\Alvl}{c_2}}{\HasType{\Alvl}{\Seqg{c_1}{c_2}}}\textsc{SeqTy} \and
    \inferrule{\L(x)\ \mathrm{is\ undefined}}{\HasType{\Alvl}{\Assign{A}{x}{E}}}\textsc{UAsgTy} \and
    \inferrule{\explevel{\lvl_E}{A}{E} \\ \lvl_E \leq \L(x)}{\HasType{\Alvl}{\Assign{A}{x}{E}}}\textsc{LAsgTy} \and
    \inferrule{ }{\HasType{\Alvl}{\Acquire{A}{\varl}}}\textsc{AcqTy} \and
    \inferrule{\forall \sigma.\ \annsatLoc{\sigma}{A} \implies \D(\L(E),\L(x),\sigma,\eval{\mem[\sigma]}{E},\DAssign{A}{x}{E})}{\HasType{\Alvl}{\DAssign{A}{x}{E}}}\textsc{DAsgTy}     \and
    \inferrule{ }{\HasType{\Alvl}{\Release{A}{\varl}}}\textsc{RelTy} \and
    \inferrule{\forall \sigma.\ \annsatLoc{\sigma}{A} \implies \D(\L(E),\lvl,\sigma,\eval{\mem[\sigma]}{E},\DOutput{A}{\lvl}{E})}{\HasType{\Alvl}{\DOutput{A}{\lvl}{E}}}\textsc{DOutTy}     \and
    \inferrule{\HasType{\Alvl}{c_1} \\ \HasType{\Alvl}{c_2} \\ \lnot \explevel{\Alvl}{A}{E} \implies \forall i.\ \bisimilar{\Alvl}{(i,c_1)}{(i,c_2)}}{\HasType{\Alvl}{\ITEg{A}{E}{\ c_1\ }{\ c_2\ }}}\textsc{IfTy} \and
    \inferrule{\lvl \leq \Alvl \implies \explevel{\Alvl}{A}{E}}{\HasType{\Alvl}{\Output{A}{\lvl}{E}}}\textsc{OutTy} \and
    \inferrule{\L(x)\ \mathrm{is\ undefined}}{\HasType{\Alvl}{\Input{A}{x}{\lvl}}}\textsc{UInTy} \and
    \inferrule{\lvl \leq \L(x)}{\HasType{\Alvl}{\Input{A}{x}{\lvl}}}\textsc{LInTy} \and
    \inferrule{\HasType{\Alvl}{c} \\ \explevel{\Alvl}{A}{E} \\ \explevel{\Alvl}{I}{E}}{\HasType{\Alvl}{\Whileg{A}{E}{I}{\ c}}} \textsc{WhileTy}
  \end{mathpar}
  \caption{Rules of the \OurSystem logic.\label{fig:logic}}
\end{figure*}

\subsection{Precise Reasoning with Annotations}

The rules for \OurSystem explicitly make use of the annotations~$\ann{A}$
on program commands to achieve highly precise reasoning, while still presenting
a simple logic to the user.  This is evident in the simplicity of many of
the rules of \autoref{fig:logic}. To understand how annotations are used
to achieve precise reasoning, consider the rule~\textsc{OutTy}
for outputting on channel $\lvl$. When this output is visible to the
attacker ($\lvl \leq \Alvl$), this rule uses the annotation~$A$
to reason about the sensitivity of the data contained in the expression~$E$
at this point in the program, specifically to check that this sensitivity
is no higher than the attacker level~$\Alvl$. This is captured by the
predicate~$\explevel{\lvl_A}{A}{E}$.

For a security level~$\lvl$, annotation~$A$ and
expression~$E$, $\explevel{\lvl}{A}{E}$ holds when, under~$A$,
the sensitivity of the data contained in~$E$ is not greater
than~$\lvl$. This is \emph{not} a policy statement
about~$\L(E)$ but, rather, uses~$A$ to over-approximates~$E$'s
sensitivity at this
point in the program.
\[
\begin{array}{@{}l}
  \explevel{\lvl}{A}{E} \equiv \forall \sigma\ \sigma'.\ \annsatLoc{\sigma}{A} \mathrel{\land} \annsatLoc{\sigma'}{A} \mathrel{\land} \gequiv{\lvl}{\sigma}{\sigma'}\\
  \;\; \implies \eval{\mem[\sigma]}{E} = \eval{\mem[\sigma']}{E}
\end{array}
\]

Similarly, the declassification rules \textsc{DAsgTy} and~\textsc{DOutTy}
use the annotation~$A$ to reason precisely about whether $\D$
holds at this point during the program.

\subsection{Secret-Dependent Branching}\label{sec:secret-branching}

The rule \textsc{WhileTy} for loops does not allow the loop guard to depend
on a secret. In contast, the rule \textsc{IfTy} for reasoning
about conditionals ``${\ITEg{A}{E}{\ c_1\ }{\ c_2\ }}$''does.
Its final premise
appplies when
the sensitivity of the
condition~$E$ exceeds that which can be observed by the
attacker~$\Alvl$, i.e.\ when the if-condition has branched on a secret
that should not be revealed to the attacker.
The premise
requires proving
that the two branches~$c_1$ and~$c_2$ are
$\Alvl$-\emph{bisimilar}, i.e.\ that the attacker cannot distinguish the execution of~$c_1$ from
the execution of~$c_2$.
The formal definition of bisimilarity
\ifExtended
(\autoref{defn:bisimilar}) appears in
the appendix.
\fi

\OurSystem
includes a set of proof rules to determine
whether two commands are $\Alvl$-bisimiar. These rules have been proved sound but,
due to lack of space,
we refer the reader to our Isabelle formalisation for the full details.
Briefly, these rules check that both commands
(1)~perform
the same number of execution steps,  (2)~modify no labelled variables~$x$
for which~$\L(x) \leq \Alvl$, (3)~never input from or output to 
channels~$\lvl \leq \Alvl$, 
and (4)~perform no declassifications. Thus
the $\Alvl$-attacker cannot tell which command was executed, including via scheduling effects.

One is of course free to implement other analyses to determine
bisimilarity.
Hence, \OurSystem provides a modular interface for
reasoning about secret-dependent branching.

\subsection{Soundness}\label{sec:soundness}

Recall that the soundness theorem requires the concurrent program
(with initial thread states)~$\ls[] = (\ls[0],\ldots,\ls[n-1])$ to
satisfy all of its functional correctness annotations. When this is the
case we write $\annsat \ls[]$.

\begin{theorem}[Soundness]\label{thm:soundness}
  Let $\ls[] = ((0,c_0),\ldots,(n-1,c_{n-1}))$ be the initial local thread
  states of a concurrent program.
  If $\annsat \ls[]$ holds and 
  $\HasType{\Alvl}{c_i}$ holds for all $0 \leq i < n$, then
  the program satisfies system security, i.e.\ 
  $\syssecure{\Alvl}{\ls[]}$.
\end{theorem}

In practice one applies the \OurSystem logic for an arbitrary attacker
security level~$\Alvl$, meaning that system security will hold for attackers
at all security levels.

The condition $\annsat \ls[]$ can be discharged using any of the
techniques implemented in \OurSystem (see \autoref{sec:concurrent-functional-correctness}; Step~\circled{3}),
or via any sound correctness verification method.


\section{Further Examples}\label{sec:examples}

\subsection{The Example of \autoref{fig:example}}
\label{sec:example-with-toggle}
Recall that the concurrent program of  \autoref{fig:example} implements
an extensional delimited release style policy~$\E$ defined in \autoref{defn:example-dr} \ifExtended (see also \autoref{defn:example-dr-extensional} in the appendix) \fi.

We add a fifth thread, which toggles $\inmode$ and $\outmode$
while ensuring they agree, and sets~$\valid$ to zero.

\noindent\begin{examplecode}
  \Seq{\Acquire{\nextann}{$\varl$}}
      {\Seq{\Assign{\nextann}{$\valid$}{0}}
           {\Seq{\Assign{\nextann}{$\inmode$}{$\inmode + 1$}}
                {\Seq{\Assign{\nextann}{$\outmode$}{$\inmode$}}
                     {\Release{\nextann}{$\varl$}}
                }
           }
      }
\end{examplecode}

Proving that this 5-thread
program~$\ls[]$ satisfies this policy is relatively straightforward
using \OurSystem. We employ \OurSystem's Owicki-Gries implementation
to prove that
it satisfies its annotations: $\annsat{\ls[]}$.
We then use the delimited release encoding (\autoref{defn:dr-encoding})
to generate the \OurSystem declassification policy~$\D$ that encodes
the delimited release policy.
Next, we use the rules of the \OurSystem logic to
compositionally prove that each thread~$\ls[i]$
is secure for an arbitrary security level~$\lvl$: $\HasType{\lvl}{\ls[i]}$.
From this proof, since we never use the part of the \textsc{IfTy} rule
for branching on secrets, it follows that the program is
free of $\E$-secret branching (we prove this result in general in our
Isabelle formalisation).
Then, by the soundness theorem (\autoref{thm:soundness})
the program satisfies \OurSystem's system security property
$\syssecure{\lvl}{\ls[]}$ for arbitrary~$\lvl$.
Finally, by the delimited release embedding theorem (\autoref{thm:dr-embedding}) it satisfies its delimited release policy~$\E$.

\subsection{Confirmed Declassification}\label{sec:confirmed-declass}

Besides delimited release-style policies, \OurSystem is geared to verifying
state-dependent declassification policies. Such policies are common in
systems in which interactions with trusted users authorise declassification
decisions. For example, in a sandboxed desktop operating system like
Qubes~OS~\cite{Rutkowska_Wojtczuk_10}, a user can copy sensitive files
from a protected domain into a less protected one, via an
explicit dialogue that requires the user to \emph{confirm} the
release of the sensitive information. Indeed, user interactions to make
explicit (e.g. ``Application X wants permission to access your microphone\ldots'')
or implicit~\cite{Roesner_KMPWC_12} information
access decisions are common in modern
computer systems. Yet verifying that concurrent programs only
allow information access after successful user confirmation has remained
out of reach for prior logics. We show how \OurSystem supports such
by considering a modification to \autoref{fig:example}.

Specifically, suppose the thread in \autoref{fig:example-demux} is replaced
by the one in \autoref{fig:example-demux-confirm}~(left). Instead of using the
signature check function~$\CHECKSIG$ to decide whether to declassify the
$\top$ input, it now asks the user by first outputting the value to be
declassified on channel~$\top$ and then receiving from the user a boolean
response on channel~$\bot$. 

Naturally the user is trusted, so it is appropriate for their response to
this question to
be received on the $\bot$ channel. Recall that~$\bot$ here means
that the information has minimal secrecy, not minimal integrity. Also,
since the threat model of \autoref{sec:threat-model}
forbids the attacker from supplying channel inputs, we can safely trust
the integrity of the user response.

The declassification policy is then specified
(see \autoref{fig:example-demux-confirm}~right)
as a \OurSystem
runtime state-dependent
declassification predicate~$\D$. This predicate specifies that at all times,
the most recent output sent (to the user to confirm) on the~$\top$ channel is allowed to be
declassified precisely when the most recent input consumed from the~$\bot$
channel is~1.

A complete formal statement of the policy satisfied by this example
is relegated to 
\ifExtended
\autoref{defn:confirmed-declass-extensional} in the appendix,
\fi
since it is a
trivial unfolding
of \autoref{defn:sys-secure}.
The resulting property is purely extensional,
since $\D$ above refers only to the program's
input/output behaviour. 

Proving the modified concurrent program secure proceeds similarly as for
\autoref{sec:example-with-toggle}.
%
%
This example demonstrates \OurSystem's advantages over
contemporary logics like \Covern~\cite{Murray_SE_18} and \SecCSL~\cite{Ernst_Murray_19}, which cannot handle
declassification. The example mimics the
software functionality of the
Cross Domain Desktop Compositor~\cite{Beaumont_MM_16} (CDDC), to which these
logics were recently applied~\cite{Murray_SE_18,Ernst_Murray_19}, but---crucially---includes the
addition of the CDDC's
confirmed-cut-and-paste declassification  functionality, which they cannot
verify.

\begin{figure}
  \begin{tabular}{l@{\,\,\,}r}
    \begin{minipage}{0.23\textwidth}
    \begin{examplecode}
      \Seq{\Acquire{\nextann}{$\varl$}}
          {\Seq{\ITNoElse{\nextann}{$\valid = 1$}{\ITE{\nextann}{$\outmode$ = 0}
                    {\Assign{\nextann}{$\lbuf$}{$\buf$}}
                    {\Seq{\Assign{\nextann}{$\hbuf$}{$\buf$}}
                      {\Seq{\Output{\nextann}{$\top$}{$\buf$}}
                           {\Seq{\Input{\nextann}{$\useranswer$}{$\bot$}}
                                {\ITNoElse{\nextann}{$\useranswer$ = 1}
                                        {\DAssign{\nextann}{$\lbuf$}{$\hbuf$}}
                                }
                           }
                      }
                    }}
               }{\Release{\nextann}{$\varl$}}}
    \end{examplecode}\end{minipage} &
    \fbox{\begin{minipage}{0.23\textwidth}
\noindent\[
\begin{array}{@{\!\!\!}l}
  \D(\lsrc,\ldst,\sigma,v,E) = \\
  \;\; (v = \lastoutput{\top}{\sigma} \mathrel{\land} \\
  \;\;\; \lastinput{\bot}{\sigma} = 1)
\end{array}
\]
\end{minipage}}
\end{tabular}
    \caption{User-confirmed declassification.\label{fig:example-demux-confirm}}
  \end{figure}

\subsection{Running Average}


As a final stateful-declassification example,
consider \autoref{fig:example-avg}.
The top-left thread inputs~$\top$-sensitive numbers into the ($\top$-labelled)
variable~$\readbuf$
and keeps a running  sum
of the values seem so far in the
($\top$-labelled) variable $\inputsum$,
as well as counting the number
of such values consumed in the ($\bot$-labelled) variable~$\inputcount$.
The security policy allows the average of the $\top$
inputs consumed to be declassified so long as the program has consumed
more inputs than whatever \emph{threshold}
is stored in the ($\bot$-labelled) variable~$\mininputs$.
\[
\begin{array}{l}
\D(\lsrc,\ldst,\sigma,v,E) = \\
\;\;\; \IfThenElse{\length(\levelinputs{\top}{\sigma}) \geq \mem[\sigma](\mininputs)}{\\\;\;\;\;\;\;v = \avg(\levelinputs{\top}{\sigma}}{\false}
\end{array}
\]
Here the function~$\levelinputs{\lvl}{\sigma}$ extracts from $\trace[\sigma]$
all inputs consumed so far on channel~$\lvl$.

The threshold~$\mininputs$ is dynamically updated by the bottom-left
thread; the right thread
does the declassification.

Thus this system implements
a dynamic declassification policy whose enforcement requires careful coordination between
the three threads. 
\begin{figure}
  \begin{tabular}{m{4.1cm}m{4cm}}
  \begin{subfigure}[b]{4.4cm}
    \begin{examplecode}
      \Seq{\Acquire{\nextann}{$\varl\avg$}}
          {\Seq{\Input{\nextann}{$\readbuf$}{$\top$}}
               {\Seq{\Assign{\nextann}{$\inputcount$}{$\inputcount + 1$}}
                    {\Seq{\Assign{\nextann}{$\inputsum$}{$\inputsum + \readbuf$}}
                         {\Release{\nextann}{$\varl\avg$}}
                    }
               }
          }
    \end{examplecode}
    \caption{Computing a running sum.\label{fig:example-avg-sum}}
  \end{subfigure} 
  \begin{subfigure}[t]{5cm}
    \vspace{1ex}
    \begin{examplecode}
      \Seq{\Acquire{\nextann}{$\varl\mininputs$}}
          {\Seq{\Assign{\nextann}{$\mininputs$}{$\mininputs + 1$}}
               {\Release{\nextann}{$\varl\mininputs$}}
          }
    \end{examplecode}
    \caption{Increasing the minimum threshold.\label{fig:example-avg-inc}}
  \end{subfigure}
  &
  \begin{subfigure}[b]{4cm}
    \begin{examplecode}
      \Seq{\Acquire{\nextann}{$\varl\avg$}}
          {\Seq{\Acquire{\nextann}{$\varl\mininputs$}}
               {\Seq{\ITNoElse{\nextann}{$\inputcount > \mininputs$}
                         {\ITNoElse{\nextann}{$\inputcount > 0$}
                           {\DOutput{\nextann}{$\bot$}{$(\inputsum / \inputcount)$}}
                         }
                    }
                    {\Seq{\Release{\nextann}{$\varl\mininputs$}}
                         {\Release{\nextann}{$\varl\avg$}}
                    }
               }
          }
    \end{examplecode}
    \caption{Declassifying the average.\label{fig:example-avg-decl}}
  \end{subfigure}\\
  \end{tabular}
  \caption{Declassifying the average with dynamic threshold.\label{fig:example-avg}}
\end{figure}
While its proof is similar to the previous to examples, 
the declassification
policy of this example refers to internal program state
(the variable~$\mininputs$), so is not extensional, unlike the prior examples.
Our Isabelle formalisation contains a modified version of this example
that satisfies an extensional policy in which the dynamic
threshold is given by the $\bot$ inputs received.

\section{Related Work}\label{sec:related-work}

\OurSystem targets compositional and precise
verification of expressive forms of
information flow security for shared-memory concurrent programs, by
decoupled functional correctness (DFC).
Prior techniques typically trade precision for expressiveness or vice-versa,
or depart from realistic attacker models altogether~\cite{Bastys+:PLAS18}.

Two recent logics that favour precision over expresiveness are
\Covern~\cite{Murray_SE_18} and \SecCSL~\cite{Ernst_Murray_19}.
\OurSystem is more expressive than
\Covern~\cite{Murray_SE_18} and \SecCSL~\cite{Ernst_Murray_19}, because
of \OurSystem's ability to reasoning about
declassification, which these prior logics cannot handle.

The \OurSystem logic is arguably much simpler than \Covern, even while being
more expressive. Like \OurSystem, \SecCSL is also relatively simple and clean.
However, it focuses on automated application via symbolic execution~\cite{Ernst_Murray_19}.
\OurSystem, in contrast,
was designed to favour precise reasoning over automation.

\OurSystem borrows its functional correctness annotations
from the
Owicki-Gries proof technique~\cite{Owicki_Gries_76}.
In doing so, it inherits
a well known~\cite{Malkis_Mauborgne_11} property of that technique:
the need to sometimes introduce \emph{ghost variables} when reasoning about
certain concurrent programs.
Such an example is depicted in
\ifExtended
\autoref{fig:example-covern} in
the appendix, where the ghost variable $\GHOST$ 
encodes a suitable analogue of the relational invariants of \Covern and \SecCSL
via the (non-relational) \OurSystem program annotations.
\fi

Karbyshev et al.~\cite{Karbyshev+:POST18} present a highly precise
separation logic based method for compositionally proving security of
concurrent programs. Unlike \OurSystem, their
approach supports a far more flexible scheduler model,
including reasoning about benign races on public variables,
dynamic
thread creation and thread$\rightarrow$scheduler interactions.
Unlike \OurSystem, \cite{Karbyshev+:POST18} doesn't support declassification.

Others have examined information flow verification for \emph{distributed}
concurrent programs, in which threads do not share memory.
Bauerei{\ss} et al.~\cite{Bauereiss_GPR_17} present a method for verifying
the security of
such programs,
including for some declassification policies and apply it to verify
the key functionality of a distributed social media platform. Li et~al.~\cite{Li_MT_17} present
a rely-guarantee based method, tailored to systems in which the
\emph{presence} of messages on a channel can reveal sensitive information.
In \OurSystem, input is always assumed to be available on all channels.

Decoupled functional correctness was foreshadowed
in the recent work of Li and Zhang~\cite{Li_Zhang_17}
(as well as in aspects of Amtoft et~al.~\cite{Amtoft_BB_06}).
Li and Zhang's approach supports relatively precise
reasoning about data-dependent sensitivity of sequential (i.e.\ non-concurrent)
programs that carry annotations on assignment statements.
\OurSystem
extends this idea across the entire program and applies it to
compositional reasoning about shared-memory concurrent programs.



\emph{Relational decomposition}~\cite{Beringer_Hofmann_07,Beringer_11} and the \emph{product program} approaches~\cite{Darvas_HS_05,Barthe_DR_11,Terauchi_Aiken_05} encode security reasoning via
functional correctness. Instead
\OurSystem exploits
compositional functional correctness to aid security reasoning.

\section{Conclusion}\label{sec:concl}

We presented \OurSystem, the first compositional logic for proving
information flow security for shared-memory concurrent programs that supports
precise reasoning about expressive security policies. It
embodies a new approach to building such logics:
decoupled functional correctness.

As we demonstrated,
\OurSystem supports reasoning about myriad security policies,
including delimited
release-style declassification,
value-dependent sensitivity and runtime-state dependent declassification,
and their co-operative enforcement via non-trivial thread interactions.
\OurSystem sets a new standard for reasoning methods for concurrent information
flow security.


\ifFinal
\vspace{.1cm}
\noindent\emph{Acknowledgements}
This research was sponsored by the Department of the Navy, Office of
Naval Research, under award \#N62909-18-1-2049.  Any opinions,
findings, and conclusions or recommendations expressed in this
material are those of the author(s) and do not necessarily reflect the
views of the Office of Naval Research.
This work was partly funded by the Swedish Foundation for Strategic
Research (SSF) and the Swedish Research Council
(VR).
\fi

\bibliographystyle{IEEEtran}

\bibliography{IEEEabrv,references}

\ifExtended
\appendix

\subsection{Ancillary Definitions}\label{app:defns}

\begin{figure*}
\begin{mathpar}
  \inferrule{\lstep{(\ls[i],\sigma)}{(\ls[i]',\sigma')}}{\gstep{(\ls[0],\ldots,\ls[i],\ldots,\ls[n-1],\sigma,i\cdot\sched')}{(\ls[0],\ldots,\ls[i]',\ldots,\ls[n-1],\sigma',\sched')}}\textsc{GStep} \and
  \inferrule{\nexists y.\ \lstep{(\ls[i],\sigma)}{y}}{\gstep{(\ls[0],\ldots,\ls[i],\ldots,\ls[n-1],\sigma,i\cdot\sched')}{(\ls[0],\ldots,\ls[i],\ldots,\ls[n-1],\sigma,\sched')}}\textsc{GWait}
\end{mathpar}
\caption{Concurrent execution. Here, $\lstep{\cdot}{\cdot}$ is the small-step semantics of individual thread programs (see \autoref{fig:semantics}).\label{fig:global-exec}}
\end{figure*}

\begin{definition}[State Agreement under $\E$ (full definition)]\label{defn:dr-equiv-full}
  We say that two states~$\sigma$ and~$\sigma'$ agree under
  the delimited release policy~$\E$ for an attacker at level~$\Alvl$,
  written $\gdrequiv{\Alvl}{\E}{\sigma}{\sigma'}$, iff
  (1)~$\gequiv{\Alvl}{\sigma}{\sigma}$,
  (2)~their memories
  agree on all variables, (3)~the same number of inputs has been
  consumed so far in each, and (4) the environment obtained by
  appending the inputs consumed so far in $\sigma$ 
  to $\env[\sigma]$ agrees under $\E$ with the environment obtained by
  doing likewise to~$\sigma'$.
\end{definition}
Here, condition~(2) is encodes the simplifying assumption that the initial
memories contain no secrets. Conditions~(3) and~(4) are more complicated
than might be expected due to having generalised over all~$\sigma$: for
initial states~$\sigma$ and~$\sigma'$ in which no events have been been
performed (i.e.\ $\trace[\sigma]$ and $\trace[\sigma']$ are both empty),
condition~(3) holds trivially and condition~(4) collapses to
$\envdrequiv{\Alvl}{\E}{\env[\sigma]}{\env[\sigma']}$: agreement of the two
environments under~$\E$. In this way this more general definition is
morally
equivalent to the simpler one (\autoref{defn:dr-equiv}) of \autoref{sec:dr}.

\begin{definition}[Absence of $\E$-Secret Branching]\label{defn:no-secret-branching}
  We say that a program~$\ls[]$ \emph{doesn't branch on secrets that the delimited
  release policy~$\E$ forbids from releasing}, when observed by an attacker
  at level~$\Alvl$, when if for all schedules~$\sched$ and initial
  states~$\sigma$, if the program executes to some configuration~$y$,
  then that execution can be matched from any other initial state~$\sigma'$
  for which $\gdrequiv{\Alvl}{\E}{\sigma}{\sigma'}$ to reach
  a configuration~$y'$ whose thread local states~$\ls[y']$ is equal to~$\ls[y]$, the thread local states of~$y$ (meaning that the two runs are still
  executing the same code in all threads) and, moreover, the same
  number of $\Alvl$-visible events have been performed so far
  in~$y$ and~$y'$ and, for all levels~$\lvl$, the same number of inputs
  from channel~$\lvl$ has been consumed in both~$y$ and~$y'$.
\end{definition}

\begin{figure*}
  \begin{mathpar}
    \inferrule{\eval{\mem[\sigma]}{E} = v \\ \mem[]' = \mem[\sigma][x \mapsto v]}{\lstep{((i,\Assign{A}{x}{E}),\gstate{\sigma})}{((i,\Skip),{(\env[\sigma],\mem[]',\locks[\sigma],\trace[\sigma])})}}\textsc{Assign} \\
    \inferrule{\eval{\mem[\sigma]}{E} = v \\ \mem[]' = \mem[\sigma][x \mapsto v] \\ \trace[]' = \trace[\sigma] \cdot \devent{\L(x)}{v}{E}}{\lstep{((i,\DAssign{A}{x}{E}),\gstate{\sigma})}{((i,\Skip),{(\env[\sigma],\mem[]',\locks[\sigma],\trace[]')})}}\textsc{DAssign} \\
    \inferrule{\eval{\mem[\sigma]}{E} = v \\ \trace[]' = \trace[\sigma] \cdot \oevent{\lvl}{v}{E}}{\lstep{((i,\Output{A}{\lvl}{E}),\gstate{\sigma})}{((i,\Skip),{(\env[\sigma],\mem[\sigma],\locks[\sigma],\trace[]')})}}\textsc{Output}    \\
    \inferrule{\eval{\mem[\sigma]}{E} = v \\ \trace[]' = \trace[\sigma] \cdot \devent{\lvl}{v}{E}}{\lstep{((i,\DOutput{A}{\lvl}{E}),\gstate{\sigma})}{((i,\Skip),{(\env[\sigma],\mem[\sigma],\locks[\sigma],\trace[]')})}}\textsc{DOutput}    \\
    \inferrule{\env[\sigma](\lvl) = v \cdot \vs \\ \env[]' = \env[\sigma][\lvl \mapsto \vs] \\ \mem[]' = \mem[\sigma][x \mapsto v] \\ \trace[]' = \trace[\sigma] \cdot \ievent{\lvl}{v}}{\lstep{((i,\Input{A}{\lvl}{v}),\gstate{\sigma})}{((i,\Skip),{(\env[]',\mem[]',\locks[\sigma],\trace[]')})}}\textsc{Input}    \\
    \inferrule{\eval{\mem[\sigma]}{E} = \true}{\lstep{((i,\ITEg{A}{E}{\ c_1\ }{\ c_2\ }),\sigma)}{((i,c_1),\sigma)}}\textsc{IfT}   \and
    \inferrule{\eval{\mem[\sigma]}{E} \not= \true}{\lstep{((i,\ITEg{A}{E}{\ c_1\ }{\ c_2\ }),\sigma)}{((i,c_2),\sigma)}}\textsc{IfF}    \\
    \inferrule{\lstep{((i,c_1),\sigma)}{((i,c_1'),\sigma')} \\ c_1' \not= \Skip}{\lstep{((i,\Seqg{c_1}{c_2}),\sigma)}{((i,\Seq{c_1'}{c_2}),\sigma')}}\textsc{Seq}    \and
    \inferrule{\lstep{((i,c_1),\sigma)}{((i,c_1'),\sigma')} \\ c_1' = \Skip}{\lstep{((i,\Seqg{c_1}{c_2}),\sigma)}{((i,c_2),\sigma')}}\textsc{SeqStop}    \\       \inferrule{\eval{\mem[\sigma]}{E} = \true}{\lstep{((i,\Whileg{A}{E}{I}{\ c}),\sigma)}{((i,\Seq{c}{\Whileg{A}{E}{I}{\ c}}),\sigma)}}\textsc{WhileT}    \and
    \inferrule{\eval{\mem[\sigma]}{E} \not= \true}{\lstep{((i,\Whileg{A}{E}{I}{\ c}),\sigma)}{((i,\Skip),\sigma)}}\textsc{WhileF}    \\
    \inferrule{\locks[\sigma](\varl)\ \mathrm{is\ undefined} \\ \locks[]' = \locks[\sigma][\varl \mapsto i]}{\lstep{((i,\Acquire{A}{\varl}),\gstate{\sigma})}{((i,\Skip),(\env[\sigma],\mem[\sigma],\locks[]',\trace[\sigma]))}}\textsc{Acquire}    \\
    \inferrule{\locks[\sigma](\varl) = i \\ \locks[]' = \locks[\sigma][\varl\ \mathrm{is\ undefined}]}{\lstep{((i,\Release{A}{\varl}),\gstate{\sigma})}{((i,\Skip),(\env[\sigma],\mem[\sigma],\locks[]',\trace[\sigma]))}}\textsc{Release}    \\        
  \end{mathpar}
  \caption{Semantics of threads, where $\sigma = \gstate{\sigma}$.\label{fig:semantics}}
\end{figure*}

\subsubsection*{Extensional Policies}

\begin{definition}[Extensional Delimited Release Policy for \autoref{fig:example} and \autoref{sec:example-with-toggle}]\label{defn:example-dr-extensional}
  For completeness, we specify the extensional security property that the
  delimited release policy for \autoref{fig:example} (see \autoref{defn:example-dr}) encodes. 
\begin{DIFnomarkup}
\begin{align*}
  &\forall \sigma\ \sched\ y\ \sigma'.\;\\
  &(\forall n.\; \forall h\!\!\in\!\!\{\lambda \vs.\ \IfThenElse{\length(\vs) \not= 0}{\CHECKSIG(\last(\vs))}{0}, \\
  &\qquad\quad\quad\; \lambda \vs.\ \IfThenElse{\length(\vs) \not= 0 \mathrel{\land} \CHECKSIG(\last(\vs)) = 0\\&\quad\qquad\qquad\quad\quad\;\;\;}{
  \last(\vs)}{0} \}.\\
  &\qquad h(\take(n, \env[\sigma])) = h(\take(n, \env[\sigma']))) \mathrel{\land} \\
  & \gequiv{\bot}{\sigma}{\sigma'} \mathrel{\land}
  \gstepstar{(\ls, \sigma, \sched)}{y} \Rightarrow\\
  &\exists y'.\; \gstepstar{(\ls, \sigma', \sched)}{y'} \mathrel{\land}
   \trequiv{\bot}{\trace[y]}{\trace[y']}
\end{align*}
\end{DIFnomarkup}
  
\end{definition}

\begin{definition}[Extensional Confirmed Declassification Policy for \autoref{sec:confirmed-declass}]\label{defn:confirmed-declass-extensional}
\begin{align*}
  &\forall \sigma\ \sched\ \sigma'\ \ls'\ \sched'\ \ls''\ \sigma''\ \sched''\ e. \\
  &\gstepstar{(\ls, \sigma, \sched)}{(\ls', \sigma', \sched')} \mathrel{\land}\\
  &\gstepstar{(\ls', \sigma', \sched')}{(\ls'', \sigma'', \sched'')}
  \mathrel{\land} \trace[\sigma''] = \trace[\sigma'] . e \Rightarrow\\
  &\begin{cases}
    \begin{array}{l}
      v = \lastoutput{\top}{\sigma'} \mathrel{\land}\\
      \;\;\lastinput{\bot}{\sigma'} = 1
    \end{array} & \mathrm{if\ } e = \devent{\bot}{v}{E} \\
    \begin{array}{l}
    \uncertainty{\Alvl}{\ls[]}{\sigma}{\sched}{\trace[\sigma']} \subseteq \\
    \;\;\uncertainty{\Alvl}{\ls[]}{\sigma}{\sched}{\trace[\sigma'] \cdot e}
      \end{array} &
    \mathrm{otherwise}
  \end{cases}
\end{align*}
\end{definition}

\subsubsection*{$\Alvl$-Bisimilarity}

$\Alvl$-bisimilarity is defined via the notion of an
\emph{$\Alvl$-secure bisimulation}.
Essentially a $\Alvl$-secure bisimulation is a relational invariant on
the execution of a thread that ensures that each step of its execution
satisfies what we call \emph{$\Alvl$-step security}.

\begin{definition}[$\lvl$-Step Security]
  Let $\lvl$ be a security level. Let
  $\sigma$ and $\sigma_2$ be global states such that a single execution
  step has occurred from~$\sigma$ to reach~$\sigma_2$, and let
  $\sigma'$ and~$\sigma_2'$ be likewise, such that
  $\gequiv{\lvl}{\sigma}{\sigma'}$.
  Then these states satisfy
  $\lvl$-step security, written $\stepsecure{\lvl}{\sigma}{\sigma_2}{\sigma'}{\sigma_2'}$,
  iff:
  \begin{itemize}
  \item If the execution step from~$\sigma$ produced a declassification event visible
    at level~$\lvl$,
    then, whenever the same event is produced by the step from~$\sigma'$,
    we require that $\gequiv{\lvl}{\sigma_2}{\sigma_2'}$.
  \item If the execution step from~$\sigma$ produced a declassification event not visible at level~$\lvl$, then we require that $\gequiv{\lvl}{\sigma_2}{\sigma_2'}$ unconditionally.
  \item In either case, the number of $\lvl$-visible events in
    $\trace[\sigma_2]$ and $\trace[\sigma_2']$ must be equal.
  \item Otherwise, if no declassification event is produced in the step from~$\sigma$, we require that $\gequiv{\lvl}{\sigma_2}{\sigma_2'}$.
  \end{itemize}
\end{definition}

\begin{definition}[$\lvl$-Secure Bisimulation]
  For a security level~$\lvl$, a binary relation~$\R$ on thread local
  states~$(i,c)$ is an
  $\lvl$-secure bisimulation iff whenever $(i,c) \mathrel{\R} (i',c')$:
  \begin{itemize}
  \item $i = i'$
  \item $c = \Skip \iff c' = \Skip$
  \item An execution step of~$\lstep{((i,c),\sigma)}{((i,c_2),\sigma_2)}$
    from a global state~$\sigma$
    that satisfies~$c$'s annotation, can be matched by a step
    $\lstep{((i,c'),\sigma')}{((i,c_2'),\sigma_2')}$ from any
    global state~$\sigma'$ that satisfies~$c'$'s annotation whenever
    $\gequiv{\lvl}{\sigma}{\sigma'}$. Moreover, in that case
    $(i,c_2) \mathrel{\R} (i,c_2')$ is preserved and
    $\lvl$-step security is satisfied: $\stepsecure{\lvl}{\sigma}{\sigma_2}{\sigma'}{\sigma_2'}$.
  \end{itemize}
\end{definition}

\begin{definition}[$\lvl$-Bisimilarity]\label{defn:bisimilar}
  We say that two local thread states~$(i,c)$ and $(i',c')$ are
  $\lvl$-bisimilar, written $\bisimilar{\lvl}{(i,c)}{(i',c')}$ whenever
  there exists a $\lvl$-secure bisimulation $\R$ that relates them:
  $(i,c) \mathrel{\R} (i',c')$.
\end{definition}

\subsection{A System Requiring Ghost Variables}\label{app:covern-system}

\begin{figure*}
  \begin{center}
    \begin{tabular}{m{\figcellwidth}m{\figcellwidth}m{\figcellwidth}}
      \begin{tabular}{c}
        \begin{subfigure}[b]{\subfigslength}
          \begin{examplecode}
  \Seq{\Acquire{\nextann}{$\varl$}}
      {\Seq{\Seq{\Assign{\nextann}{$\inmode$}{$\inmode + 1$}}
                {\Seq{\Assign{\nextann}{$\outmode$}{$\inmode$}}
                     {\Release{\nextann}{$\varl$}}
                }
        }
        {\Assign{\nextann}{$\GHOST$}{0}}
      }
\end{examplecode}
    \caption{Toggling thread.\label{fig:example-covern-toggle}}
      \end{subfigure}
      \bigskip \\
  \begin{subfigure}[b]{\subfigslength}
    \begin{examplecode}
      \Output{\nextann}{$\top$}{$\hbuf$}
    \end{examplecode}
    \caption{Outputting $\top$ data.\label{fig:example-covern-top}}
  \end{subfigure}
  \end{tabular}
      &
      \begin{tabular}{l}
      \begin{subfigure}[b]{\subfigslength}
    \begin{examplecode}
      \Seq{\Acquire{\nextann}{$\varl$}}{\Seq{\ITE{\nextann}{$\inmode$ = 0}{\Input{\nextann}{$\buf$}{$\bot$}}{\Input{\nextann}{$\buf$}{$\top$}}}}{\Seq{\Assign{\nextann}{$\GHOST$}{1}}{\Release{\nextann}{$\varl$}}}
    \end{examplecode}
    \caption{Reading data into a shared buffer.\label{fig:example-covern-mux}}
      \end{subfigure}
      \bigskip \\
  \begin{subfigure}[b]{\subfigslength}
    \begin{examplecode}
      \Output{\nextann}{$\bot$}{$\lbuf$}
    \end{examplecode}
    \caption{Outputting $\bot$ data.\label{fig:example-covern-bot}}
  \end{subfigure}
  \end{tabular}      
      &
  \begin{subfigure}[b]{\subfigslength}
    \begin{examplecode}
      \Seq{\Acquire{\nextann}{$\varl$}}
          {\Seq{\ITE{\nextann}{$\outmode$ = 0}
                    {\Assign{\nextann}{$\lbuf$}{$\buf$}}
                    {\Assign{\nextann}{$\hbuf$}{$\buf$}}
               }{\Release{\nextann}{$\varl$}}}
    \end{examplecode}
    \caption{Copying data from a shared buffer.\label{fig:example-covern-demux}}
  \end{subfigure}
    \end{tabular}
    \caption{Co-operative Use of a Shared Buffer without declassification\label{fig:example-covern}. The variable $\GHOST$ is a \emph{ghost variable} needed to verify this system in \OurSystem.
    Specifically, it distinguishes the cases in which $\buf$ is newly cleared
by the toggle thread (\autoref{fig:example-covern-toggle})
and those in which the reading thread (\autoref{fig:example-covern-mux})
has overwritten~$\buf$: these two cases do not need to be distinguished
when using a relational invariant~\cite{Murray_SE_18,Ernst_Murray_19}
to describe~$\buf$'s (value-dependent)
sensitivity in \Covern, but this distinction is required when encoding this same information via the \OurSystem annotations.
    }

  \end{center}
\end{figure*}

\fi 

\end{document}